\documentclass[11pt,epsf,epsfig]{article}
\usepackage{graphics}
\usepackage{graphicx}
\usepackage{amssymb}
\usepackage{amsmath}

\newcommand{\gsim}{\lower.7ex\hbox{$\;\stackrel{\textstyle>}{\sim}\;$}}
\newcommand{\lsim}{\lower.7ex\hbox{$\;\stackrel{\textstyle<}{\sim}\;$}}

\def\stilde{\widetilde}

\newcommand{\newc}{\newcommand}
\newc{\Nc}{N_{c}}
\newc{\CG}{C_G}
\newc{\gp}{g'}
\newc{\stopi}{\stilde t_i}
\newc{\sboti}{\stilde b_i}
\newc{\staui}{\stilde \tau_i}
\newc{\stopj}{\stilde t_j}
\newc{\sbotj}{\stilde b_j}
\newc{\stauj}{\stilde \tau_j}
\newc{\stopI}{\stilde t_1}
\newc{\stopII}{\stilde t_2}
\newc{\sbotI}{\stilde b_1}
\newc{\sbotII}{\stilde b_2}
\newc{\stauI}{\stilde \tau_1}
\newc{\stauII}{\stilde \tau_2}
\newc{\sstop}{s_{t}}
\newc{\cstop}{c_{t}}
\newc{\ssbot}{s_{b}}
\newc{\csbot}{c_{b}}
\newc{\sstau}{s_{\tau}}
\newc{\cstau}{c_{\tau}}
\newc{\Sstop}{s_{2t}}
\newc{\Cstop}{c_{2t}}
\newc{\Ssbot}{s_{2b}}
\newc{\Csbot}{c_{2b}}
\newc{\Sstau}{s_{2\tau}}
\newc{\Cstau}{c_{2\tau}}
\newc{\salpha}{s_\alpha}
\newc{\calpha}{c_\alpha}
\newc{\Calpha}{c_{2\alpha}}
\newc{\Salpha}{s_{2\alpha}}
\newc{\sbetapm}{s_{\beta_\pm}}
\newc{\cbetapm}{c_{\beta_\pm}}
\newc{\Sbetapm}{s_{2 \beta_\pm}}
\newc{\Cbetapm}{c_{2 \beta_\pm}}
\newc{\sbetaO}{s_{\beta_0}}
\newc{\cbetaO}{c_{\beta_0}}
\newc{\SbetaO}{s_{2 \beta_0}}
\newc{\CbetaO}{c_{2 \beta_0}}
\newc{\vu}{v_u}
\newc{\vd}{v_d}
\newc{\seL}{\stilde e_L}
\newc{\smuL}{\stilde \mu_L}
\newc{\seR}{\stilde e_R}
\newc{\smuR}{\stilde \mu_R}
\newc{\suL}{\stilde u_L}
\newc{\sdL}{\stilde d_L}
\newc{\suR}{\stilde u_R}
\newc{\sdR}{\stilde d_R}
\newc{\scL}{\stilde c_L}
\newc{\ssL}{\stilde s_L}
\newc{\scR}{\stilde c_R}
\newc{\ssR}{\stilde s_R}
\newc{\snue}{\stilde \nu_e}
\newc{\snumu}{\stilde \nu_\mu}
\newc{\snutau}{\stilde \nu_\tau}
\newc{\Gpm}{G^\pm}
\newc{\Hpm}{H^\pm}
\newc{\FFbS}{\overline{FF}S}
\newc{\FFbV}{\overline{FF}V}
\newc{\FSS}{F_{SS}}
\newc{\FSSS}{F_{SSS}}
\newc{\FFFS}{F_{FFS}}
\newc{\FFFbS}{F_{\overline{FF}S}}
\newc{\FSSV}{F_{SSV}}
\newc{\FVS}{F_{VS}}
\newc{\FVVS}{F_{VVS}}
\newc{\FFFV}{F_{FFV}}
\newc{\FFFbV}{F_{\overline{FF}V}}
\newc{\Fgauge}{F_{\rm gauge}}
\newc{\DRbarprime}{$\overline{\rm DR}'$ }
\newc{\DRbar}{$\overline{\rm DR}$ }
\newc{\MSbar}{$\overline{\rm MS}$ }
\newc{\Yu}{{\bf Y}_u}
\newc{\Yd}{{\bf Y}_d}
\newc{\Ye}{{\bf Y}_e}
\newc{\Au}{{\bf a}_u}
\newc{\Ad}{{\bf a}_d}
\newc{\Ae}{{\bf a}_e}
\newc{\bm}{{\bf m}}

\newc{\rwino}{r_{\tilde W}}
\newc{\rmu}{r_{\tilde H}}
\newc{\ra}{r_A}

\newcommand{\nc}{\newcommand}

\nc{\beaa}{\begin{eqnarray*}} \nc{\eeaa}{\end{eqnarray*}}
\nc{\beq}{\begin{equation}}   \nc{\eeq}{\end{equation}}
\nc{\bea}{\begin{eqnarray}}   \nc{\eea}{\end{eqnarray}}
\nc{\baa}{\begin{array}}      \nc{\eaa}{\end{array}}
\nc{\bit}{\begin{itemize}}    \nc{\eit}{\end{itemize}}
\nc{\ben}{\begin{enumerate}}  \nc{\een}{\end{enumerate}}
\nc{\bce}{\begin{center}}     \nc{\ece}{\end{center}}
\nc{\non}{\nonumber}

\def\bed{\begin{description}}
\def\eed{\end{description}}

\def\non{\nonumber}

\def\k1slash{k_1\hspace{-10.5pt}/\ \ }

\def\phiphi{\phi_{{}_\Phi}}

\def\phiphibar{\phi_{{}_{\overline{\Phi}}}}

\def\phichi{\phi_{{}_\chi}}

\def\simge{\mathrel{%
   \rlap{\raise .57ex \hbox{$>$}}{\lower .57ex \hbox{$\sim$}}}}
\def\simle{\mathrel{
   \rlap{\raise 0.512ex \hbox{$<$}}{\lower 0.512ex \hbox{$\sim$}}}}

\begin{document}

\title{\textbf{Phenomenology of Dirac Neutrinogenesis}\\
\textbf{in Split Supersymmetry}}
\author{Brooks Thomas and Manuel Toharia \\
\it{Michigan Center for Theoretical Physics (MCTP)} \\
\it{Department of Physics, University of Michigan, Ann Arbor, MI 48109}}
\date{November 16, 2005}
\maketitle

\begin{abstract}

In Split Supersymmetry scenarios the possibility of having a very
heavy gravitino opens the door to alleviate or completely solve the
worrisome ``gravitino problem'' in the context of supersymmetric
baryogenesis models.
Here we assume that the gravitino may indeed be
heavy and that Majorana masses for neutrinos are forbidden as well
as direct Higgs Yukawa couplings between left and right handed
neutrinos.
We investigate the viability of the mechansim known as Dirac
leptogenesis (or neutrinogenesis), both in solving the baryogenesis
puzzle and explaining the observed neutrino sector phenomenology.
To successfully address these issues, the scenario requires the
introduction of at least two new heavy fields.  If a hierarchy among
these new fields is introduced, and some reasonable stipulations are
made on the couplings that appear in the superpotential, it becomes
a generic feature to obtain the observed large lepton mixing angles.
We show that in this case, it is possible simultaneously to obtain
both the correct neutrino phenomenology and enough baryon number,
making thermal Dirac neutrinogenesis viable. However, due to
cosmological constraints, its ability to satisfy these constraints
depends nontrivially on model parameters of the overall theory,
particularly the gravitino mass.  Split supersymmetry with
\(10^{5}\)~GeV~\(\lesssim m_{3/2} \lesssim 10^{10}\)~GeV emerges as
a ``natural habitat" for thermal Dirac neutrinogenesis.

%

\end{abstract}

\section{Introduction}

\indent

The phenomenology of split supersymmetry
models~\cite{Arkani-Hamed:2004fb,Giudice:2004tc} in which a
hierarchy is generated between the gaugino masses and the masses of
the scalar sparticles has received a great deal of attention in
recent times. The main advantage of such scenarios is that they
circumvent a wide variety of data pressures on theories with
supersymmetry breaking on a lower scale that arise from potentially
dangerous radiative corrections involving TeV-scale scalar
sparticles for the Higgs Mass, flavor-violating effects, etc.  This
is done by making the scalar masses heavy, while keeping the gaugino
masses at the TeV scale or below to constitute the dark matter and
to preserve gauge coupling unification.

\indent In one particularly simple scenario, which we refer to as
loop-split supersymmetry (after the PeV-scale supersymmetry
of~\cite{Wells:2003tf}), supersymmetry is broken at an intermediate
scale, around \(\sim10^{5}-10^{7}\) GeV, and all scalars in the
theory, (with the exception of one light Higgs particle) are given
masses around the PeV scale while the gauginos acquire masses at the
TeV scale~\cite{Randall:1998uk}.  The preferred method of achieving
this hierarchy is by invoking anomaly mediation in the gaugino
sector (but not the scalar sector).  This can be arranged by
charging the chiral supermultiplet \(X\) responsible for the
transmission of supersymmetry effects under some symmetry, to the
effect that the term that normally provides the dominant
contribution to the gaugino mass is forbidden by gauge invariance.
The dominant contribution to the gaugino masses now arises only at
the one-loop level~\cite{Randall:1998uk} and is given by
\begin{equation}
  M_{\lambda}=\frac{\beta_{g_{\lambda}}}{g_{_{\lambda}}}
  \left(\frac{\langle F_{X}^{\dagger}F_{X}\rangle}{M_{P}^{2}}\right)^{1/2},
  \label{eq:AMSBMass}
\end{equation}
where the index \(\lambda\) labels the gauge groups in the theory.
Charging \(X\) will not affect the scalar masses, which are still
manifestly gauge-invariant whether \(X\) is charged or not, and we
obtain the desired hierarchy.  The advantage of this scenario is its
simplicity: no additional symmetries are required to keep the
gaugino masses from being evolved up to the SUSY-breaking scale. The
theory also imposes severe restrictions on both the identity
(which~(\ref{eq:AMSBMass}) dictates ought to be predominantly either
Wino or Higgsino for strict one-loop AMSB) and mass of the lightest
supersymmetric particle (LSP), and it has been shown that dark
matter composed primarily of such particles could have observable
consequences at the next generation of \(\gamma\)-ray
telescopes~\cite{Masiero:2004ft,Arvanitaki:2004df,Thomas:2005te}.
Also, in loop-split SUSY, characteristic signatures of gluino decays
may be observable at colliders~\cite{Toharia:2005gm,Gambino:2005eh}.

\indent As mentioned above, one of the reasons why split
supersymmetry is such an attractive phenomenological model is that
it provides a convenient way of circumventing data pressures on
weak-scale superpartner masses.  However, for a theory to describe
the universe we live in, it is not enough that it evade all present
experimental bounds: the theory must be cosmologically viable as
well, in the sense that it does not disrupt big bang nucleosynthesis
(BBN), that it is compatible with cosmic inflation, that it permits
some mechanism by which baryogenesis may occur, etc.  In this paper,
we examine the cosmological ramifications of split supersymmetry,
and in particular show that a viable model of baryogenesis can be
realized via Dirac neutrinogenesis (i.e. Dirac
leptogenesis)~\cite{Dick:1999je,Murayama:2002je}. This is not a
trivial problem: it turns out that constraints arising from
gravitino cosmology fix the reheating temperature associated with
cosmic inflation to \(\sim10^{10}\)~GeV or lower for substantial
regions of \(m_{3/2}-m_{\mathit{LSP}}\) parameter space.  This
heightens the tensions already inherent among the model parameters
(mass scales, couplings, etc.) of the theory. Data from neutrino
oscillation experiments impose an additional battery of constraints
any phenomenologically viable theory must satisfy. As a result,
getting Dirac neutrinogenesis to work in split supersymmetry
requires careful consideration of both cosmology and neutrino
physics.

\indent Our aim in this paper is threefold.  First, we examine the
astrophysical and cosmological constraints on thermal leptogenesis
models, especially those associated with baryogenesis and gravitino
cosmology.  Second, we investigate the neutrino spectrum constraints
imposed on Dirac neutrinogenesis models by neutrino oscillation
experiments. Third, we solve the full system of Boltzmann equations
for Dirac neutrinogenesis and assess its viability---i.e. its
ability to satisfy all aforementioned constraints.

\section{Dirac Neutrinogenesis\label{sec:DiracLep}}

\indent

The universe we live in is manifestly asymmetric between baryons and
antibaryons.  This statement can be quantified by introducing a
parameter \(\eta\), defined as \(\eta=n_{B}/n_{\gamma}\).  Here
\(n_{B}\equiv n_{b}-n_{\overline{b}}\), where \(n_{b}\) and
\(n_{\overline{b}}\) are the baryon density and antibaryon density
of our universe, respectively, and
 \(n_{\gamma}\) is the present number density of photons.  The value of \(\eta\) has recently
been measured with great precision by WMAP~\cite{Bennett:2003bz} to be within the range
\begin{equation}
  \eta=\left(6.1 \pm 0.3\right)\times 10^{-10}.
  \label{eq:WMAPeta}
\end{equation}
In the standard cosmology, which is symmetric with respect to baryons and antibaryons, one
would expect to find \(\eta=0\).  As this is not to be the case, we must find a method of
baryogenesis -- the generic term for a process through which a baryon asymmetry might evolve in
the early universe -- to account for this.  A set of
generic criteria required for any successful baryogenesis scenario were first
established by Sakharov~\cite{Sakharov:1967dj}: first, there must be baryon number \(B\)
violation; second, there must be \(C\) and \(CP\) violation; and third, there
must be some departure from thermal equilibrium.  If any one of these conditions is not met,
baryogenesis fails\footnote{Actually the third condition is only required in theories where
the Hamiltonian preserves \(CPT\).  For a theory that is not \(CPT\)-invariant, such as
the spontaneous baryogenesis of~\cite{Cohen:1987vi}, departure from thermal equilibrium is not
necessary for baryogenesis.  In the scenarios we are considering, we will assume that \(CPT\) is
a good symmetry of the Hamiltonian and that some departure from thermal equilibrium is
required.}.

\indent A variety of viable baryogenesis models exist, including
electroweak baryogenesis, in which the CP-violation occurs at a
bubble wall, or phase boundary, and Affleck-Dine
baryogenesis~\cite{Affleck:1984fy}, in which the baryon asymmetry is
generated by moduli fields charged under \(B-L\).  In this paper, we
will focus on models which achieve baryogenesis through a framework
known as leptogenesis~\cite{Fukugita:1986hr,Luty:1992un}.  In this
scenario, decays of heavy particles in the early universe which
violate both \(CP\) and lepton number \(L\) produce an initial
lepton asymmetry, which is then converted to a nonzero baryon
asymmetry by sphaleron processes associated with the SU(2)
electroweak anomaly~\cite{'tHooft:1976fv}. Leptogenesis is a
particularly attractive model because in addition to its ability to
yield a realistic value for \(\eta\)~\cite{Buchmuller:2002xm}, it
can also explain why the standard model neutrinos have small but
nonzero masses.  In its most common form, which we will call
Majorana leptogenesis, the heavy particle whose decays violate \(L\)
is taken to be the right handed neutrino, which being a gauge
singlet, may be given a large Majorana mass \(M_{\nu_{R}}\).  Since
the most general realizable superpotential
\begin{equation}
  \mathcal{W}\ni y L N H_{u} + M_{\nu_{R}}N N
  \label{eq:MajLepSuperpot}
\end{equation}
will also contain a term which gives rise to a Dirac mass
\(m_{D}=y v \sin\beta\), where \(v\) is the standard model
Higgs VEV, the neutrino mass matrix will contain small off-diagonal terms mixing
\(\nu_{L}\) and \(\nu_{R}\).  When this matrix is diagonalized, the resulting mass spectrum
contains three light neutrinos with masses
\begin{equation}
  m_{\nu}=
  m_{D}\frac{1}{M_{\nu_{R}}}m^{\dagger}_{D},
\end{equation}
which are identified with the standard model neutrinos, as well as
three heavy neutrinos with masses \(\sim M_{\nu_{R}}\).  This
see-saw mechanism~\cite{Yanagida:1979as} is an example of one of the
strongest assets of leptogenesis scenarios: they all provide some
explanation for the lightness of neutrino masses, in addition to
explaining the origin of the observed baryon asymmetry.

\begin{figure}[t!]
\hspace{1.5cm}    
\includegraphics[width=14cm]{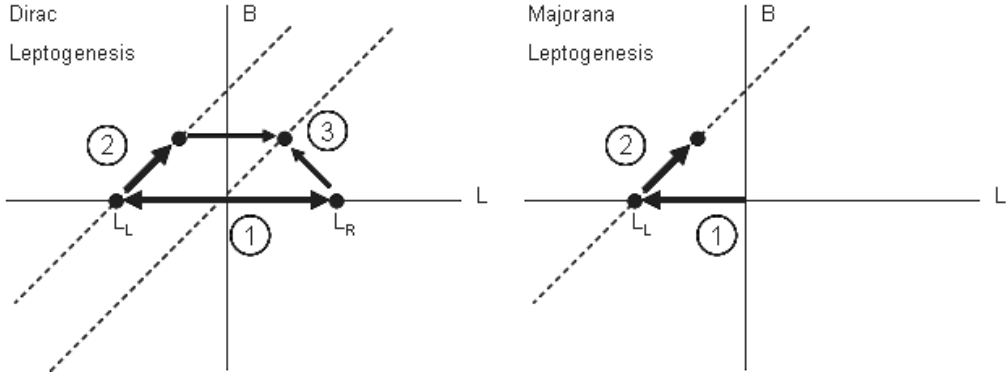}
  \caption{A schematic representation, after~\cite{Dick:1999je}, of the
  evolution of baryon number \(B\) (vertical axis) and lepton number
  \(L\) (horizontal axis) in Dirac and Majorana leptogenesis.  In
  Dirac neutrinogenesis (left panel), the evolution of \(B\) and \(L_{\mathit{tot}}\) proceeds in three steps: first,
  two stores of lepton number \(L_{\ell}\) (stored in left-handed neutrinos) and \(L_{\nu_{R}}\)
  (stored in right-handed neutrinos) are produced during heavy particle decays; second, sphaleron processes
  (which act along lines of constant \(B-L\)) mix \(L_{\ell}\) and \(B\) while leaving \(L_{\nu_{R}}\) alone; third, after
  sphaleron interactions have effectively shut off, equilibration between \(L_{\ell}\) and \(L_{\nu_{R}}\) results in a net
  positive \(B\) and \(L\) for the universe.  In Majorana leptogenesis (right-hand panel), only one store of
  lepton number is created, and the result is a universe with negative \(L_{\mathit{tot}}\) and positive \(B\).
  \label{fig:BLSpaceDiagram1}}
\end{figure}

\indent Recently, another promising leptogenesis model, which is
often called Dirac leptogenesis or Dirac
neutrinogenesis~\cite{Dick:1999je,Murayama:2002je}, has received
some attention. In this scenario, an additional symmetry is
introduced, and charges are assigned under this new symmetry in a
manner which forbids both the Majorana and Dirac mass terms
appearing in~(\ref{eq:MajLepSuperpot}) and additional heavy,
vector-like fields are introduced whose decays will violate \(CP\).
These decays build up equal and opposite lepton asymmetries
\(L_{\ell}\) and \(L_{\nu_{R}}\) in the left-handed lepton and
right-handed neutrino sectors, while maintaining an overall lepton
number for the universe
\(L_{\mathit{tot}}=L_{\ell}+L_{\nu_{R}}=0\).\footnote{Unlike in
Majorana leptogenesis, no explicit lepton-number-violating terms are
present here.} In the fermion sector these stores do not equilibrate
due to the smallness of the neutrino Dirac mass term, which only
appears in the low-energy effective theory suppressed by powers of
the mass scales associated with the heavy vector-like fields. The
electroweak sphaleron processes which convert \(L_{\ell}\) into a
baryon asymmetry \(B\) effectively shut off before \(L_{\ell}\) and
\(L_{\nu_{R}}\) have a chance to equilibrate.  The result, depicted
in fig.~\ref{fig:BLSpaceDiagram1}, is that the universe ends up with
a net positive lepton number as well as baryon number.

\indent We will consider a model with the same field content as the
one presented in~\cite{Murayama:2002je}.  In addition to the usual
quark and lepton supermultiplets (of which the left-handed lepton
multiplet \(L\) and the Higgs multiplets \(H_{u}\) and \(H_{d}\)
will be pertinent to leptogenesis), we introduce a right-handed
neutrino superfield \(N\), an exotic chiral multiplet \(\chi\), and
a number \(N_{\Phi}\) of vector-like pairs of chiral multiplets
\(\Phi_{i}\) and \(\overline{\Phi}_{i}\) (the precise value of
\(N_{\Phi}\) is unspecified: any choice of \(N_{\Phi}\geq2\) is
allowed from a baryogenesis standpoint). The charge assignments
under the additional symmetry, whatever it may be, are to be
arranged such that the most general superpotential that can be
written is\footnote{In engineering the charge assignments that lead
to such a superpotential, one must of course worry about preserving
gauge coupling unification, making sure the theory is free of gauge
theory anomalies, etc.  Since we are primarily interested in
examining the phenomenology associated with such theories, we will
not explicitly address these concerns, and appeal to high-scale
physics (e.g. additional heavy fields, the Green-Schwarz
mechanism~\cite{Green:1984sg}) to resolve these issues.}
\begin{equation}
  \mathcal{W}\ni\lambda_{i\alpha}N_{\alpha}\Phi_{i}H_{u}+
  h_{i\alpha}L_{\alpha}\overline{\Phi}_{i}\chi+
  M_{\Phi_{i}}\Phi_{i}\overline{\Phi}_{i}+\mu H_{u}H_{d},
  \label{eq:DiracLepSuperpotential}
\end{equation}
where \(\alpha\) is a family index, \(\lambda_{i\alpha}\) and
\(h_{i\alpha}\) are Yukawa couplings, \(M_{i}\) are
(supersymmetry-respecting) mass terms for the \(\Phi_{i}\) fields,
and \(\mu\) is the usual Higgs mass parameter.  The particular
symmetry introduced and charge configuration employed in arriving at
this superpotential is not of particular relevance to us, nor is it
the aim of this paper to explore the model-building possibilities
afforded by different such configurations, but an example (taken
from~\cite{Murayama:2002je}) of one that works is provided in
table~\ref{tab:U1Charges}.  \(M_{i}\), \(\lambda_{i\alpha}\), and
\(h_{i\alpha}\) may in general be complex. Both the scalar and
fermionic components of the \(\Phi\) and \(\overline{\Phi}\)
multiplets, which we denote by \(\phi\), \(\overline{\phi}\),
\(\psi_{\Phi}\), and \(\psi_{\overline{\Phi}}\) will play the role
that the \(\nu_{R}\) play in Majorana leptogenesis. As for neutrino
masses, they will appear in the low-energy effective superpotential
obtained by integrating out \(\Phi_{i}\) and
\(\overline{\Phi}_{i}\):
\begin{equation}
  \mathcal{W}_{\mathit{eff}}\ni
  \frac{\lambda_{i\alpha} h_{i\beta}}{M_{\Phi_{1}}}\chi L H_{u} N
  +\mu H_{u} H_{d}.
  \label{eq:EffDiracLepSuperpotential}
\end{equation}
If we arrange for the scalar component of \(\chi\) to acquire a VEV,
then the first term in \(\mathcal{W}_{\mathit{eff}}\) translates
into a neutrino Dirac mass
\begin{equation}
  m_{\nu\alpha\beta}= \langle\chi\rangle v \sin\beta\   \sum_{i}
  \frac{\lambda_{i\alpha} h_{i\beta}}{M_{\Phi_{i}}}.
  \label{eq:NeutMassMatrixGaugeBasis}
\end{equation}
This can be attained via an O'Raifeartaigh model of the type
employed in~\cite{Borzumati:2000mc}, in which the F-term of \(\chi\)
acquires a large VEV \(\langle F\rangle\simeq m_{3/2}M_{P}\), and
supergravity effects give rise to a nonzero VEV \(\langle
\chi\rangle\simeq16\pi m_{3/2}\kappa^{-3}\) for the scalar component
of \(\chi\), where \(\kappa\) is an undetermined dimensionless
coupling constant. Requiring that \(\langle\chi\rangle\ll
M_{\Phi_{1}}\), where \(M_{\Phi_{1}}\) denotes the lightest of the
\(M_{\Phi_{i}}\), and that the \(\lambda_{i\alpha}\) and
\(h_{i\alpha}\) are \(\mathcal{O}(1)\) or smaller, this setup yields
small yet nonzero neutrino masses without the aid of the see-saw
mechanism. Furthermore, because of the simple structure of the
neutrino mass matrix given in (\ref{eq:NeutMassMatrixGaugeBasis}),
Dirac neutrinogenesis can, under certain conditions, yield
interesting predictions about the mass hierarchy among the standard
model neutrinos.

 \begin{table}[t!]
 \begin{center}
  \begin{tabular}{ccccc}
     Field & \(U(1)_{L}\) & \(U(1)_{N}\) & \(SU(2)\) & \(U(1)_{Y}\) \\ \hline
     \(N\) & -1 & +1 & \(\mathbf{1}\) & 0 \\
     \(L\) & +1 & 0 & \(\mathbf{2}\) & \(-\frac{1}{2}\) \\
     \(H_{u}\) & 0 & 0 & \(\mathbf{2}\) & \(\frac{1}{2}\) \\
     \(H_{d}\) & 0 & 0 & \(\mathbf{2}\) & \(-\frac{1}{2}\) \\
     \(\phi\) & +1 & -1 & \(\mathbf{2}\) & \(-\frac{1}{2}\) \\
     \(\overline{\phi}\) & -1 & +1 & \(\mathbf{2}\) & \(\frac{1}{2}\) \\
     \(\chi\) & 0 & -1 & \(\mathbf{1}\) & 0
   \end{tabular}
 \end{center}
 \caption{One possible set of charge assignments, taken from~\cite{Murayama:2002je},
 that leads to the Dirac neutrinogenesis superpotential given
 in~(\ref{eq:DiracLepSuperpotential}).  Here the additional symmetry
 employed is a \(U(1)\) (which may in principle be either global or
 local).  Only the charges of the fields relevant to leptogenesis, which include the Higgs doublets
 \(H_{u}\) and \(H_{d}\), the left-handed lepton superfield \(L\),
 the right-handed neutrino superfield \(N\), the heavy fields
 \(\Phi\) and \(\overline{\Phi}\), and the additional field
 \(\chi\), have been included.  Here,  \(U(1)_L\), \(SU(2)\), and \(U(1)_{Y}\) respectively
 denote lepton number, \(SU(2)\), and \(U(1)\) hypercharge quantum numbers, and \(U(1)_N\)
 denotes the charge under the additional \(U(1)\).\label{tab:U1Charges}}
 \end{table}

Just like in Majorana Leptogenesis, some intermediate scale is
required to set the mass scale of the heavy fields responsible for
the small neutrino masses and for the lepton number generation.
In our case, we are less constrained since the neutrino masses are
set not only by the masses $M_{\Phi_{i}}$ of the heavy fields, but
also by the VEV $\langle\chi\rangle$ of the new field $\chi$. This
leaves us more freedom to choose the mass scale of the heavy fields,
but can also be thought of as a drawback since we have lost some
predictiveness. One suggestive idea would be to make use of the
$\chi$ field and the additional symmetry to generate the
supersymmetric $\mu$ term. The effective superpotential could then
look like \beq
  \mathcal{W}_{\mathit{eff}}\ni
  \frac{\lambda_{i\alpha} h_{i\beta}}{M_{\Phi_{1}}}\chi L H_{u} N
  +Y_\chi\ \chi H_{u} H_{d}.
\eeq where $Y_\chi$ is an \(\mathcal{O}(1)\) coupling constant, and
where the Higgses should now be charged under the new hidden sector
symmetry (and therefore enlarging the list of potential problems of
the model, like cancelation of anomalies, etc.). In this situation,
we have $\mu=Y_\chi \langle\chi\rangle$ and any observation or
limits on Higgsino dark matter would directly constrain the VEV
$\langle \chi\rangle$. (Split) supersymmetry might also give us some
ideas as to how to relate the heavy fields $\Phi$ to the SUSY
breaking scale. We will come back to some of these issues when we
deal in detail with neutrino phenomenology in
section~\ref{sec:NeutSpec}.

\begin{figure}[t]
  \begin{center}
\includegraphics[width=7cm]{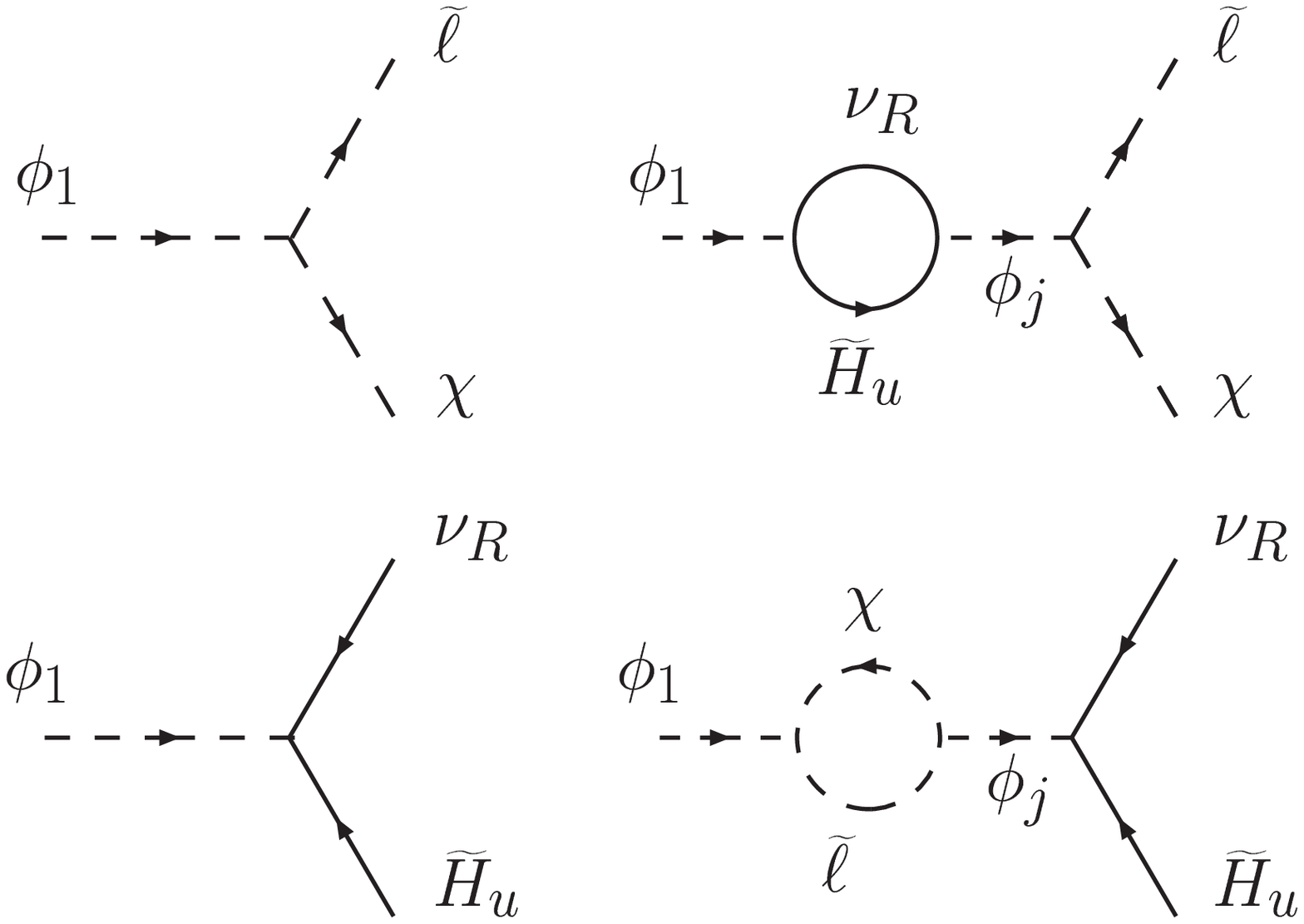}\hspace{.5cm}
\includegraphics[width=7cm]{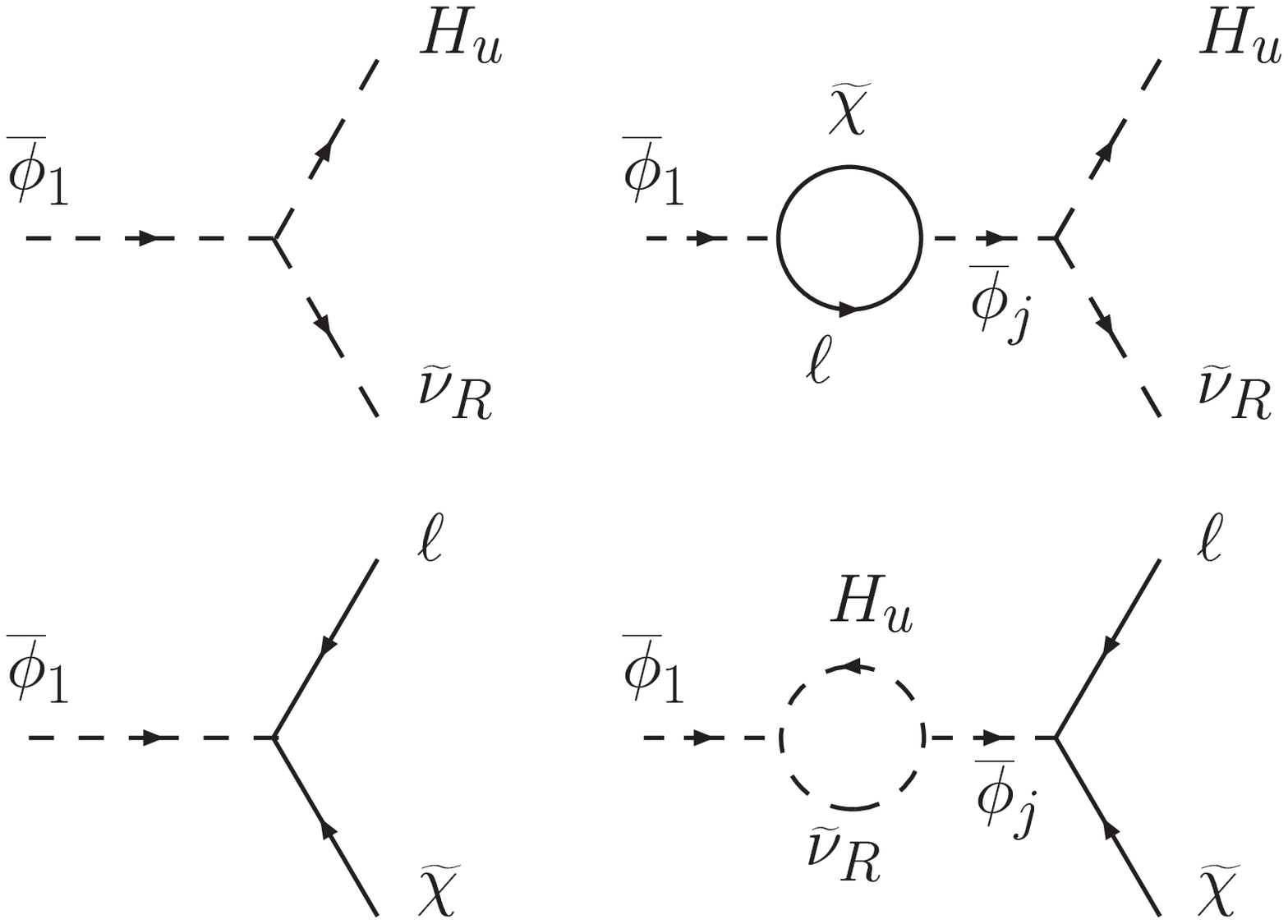}
\end{center}
\caption{Diagrams that give the leading contribution to the \(CP\) asymmetry
    from decays of the scalar fields \(\phi_{1}\) and \(\overline{\phi}_{1}\).
    Similar \(CP\) asymmetries are generated during the decay of the fermionic
    fields \(\psi_{{}_{\Phi_{1}}}\) and \(\psi_{{}_{\overline{\Phi}_{1}}}\).\label{fig:DecayDiagrams}}
\end{figure}

\indent If \(y_{i\alpha}\) and \(h_{i\alpha}\) contain nontrivial,
\(CP\)-violating phases, a net \(CP\) asymmetry will be generated as
the component fields of \(\Phi_{1}\) and \(\overline{\Phi}_{1}\)
decay, the leading contribution to which results from the
interference of tree-level and one-loop-level diagrams.  Those
relevant to \(\phi\) and \(\overline{\phi}\) decay are shown in
fig.~\ref{fig:DecayDiagrams}.\footnote{Here, we have included loop
contributions involving the scalar component of \(H_{d}\), which
start to become important when \(\mu\) is of the same order as the
\(M_{\Phi_{i}}\)}  The fermion fields \(\psi_{\Phi_{1}}\) and
\(\psi_{\overline{\Phi}_{1}}\) undergo similar decays, and stores of
lepton number are built up in the lepton fields \(\nu_{R}\) and
\(\ell\), and in the associated slepton fields \(\tilde{\nu}_{R}\),
\(\tilde{\ell}\). The \(CP\) amplitudes (and resultant
\(CP\)-asymmetries) generated from the decays of the charged fields
in the fields charged under \(SU(2)\) will be the same as those
generated from the decays of the neutral fields, since \(SU(2)\) is
unbroken. Additionally, in the approximation of unbroken
supersymmetry, the amplitudes (and resultant \(CP\)-asymmetries) in
the fermion case will be the same as those for the scalar case.  We
will assume that any lepton number stored in the slepton sector
rapidly equilibrates through the large \(\langle F_{\chi}\rangle\)
term, and thus we need only pay attention to the lepton number
stored in the lepton sector.

\indent We will begin by examining a toy model in which there are
only two sets of \(\Phi\) and \(\overline{\Phi}\), the minimum
number required for \(CP\)-violation to take place, and make the
further simplifying assumption that the masses \(M_{\Phi_{i}}\) are
all much larger than \(\mu\). In this case~\cite{Flanz:1996fb}, one
may parameterize the associated lepton number violation by defining
a single decay asymmetry \(\epsilon\), which represents the amount
of lepton number generated in any particular lepton-number-carrying
species by the decay of a single heavy particle.  This implies the
relations
\begin{eqnarray}
  \Gamma(\Phi_{1}\longrightarrow N_{\alpha}^{c}\tilde{H}^{c}_{u})-\Gamma(\Phi_{1}^{c}\longrightarrow N_{\alpha}H_{u}) & \equiv & \epsilon\Gamma_{D} \label{eq:1steps} \\
  \Gamma(\Phi_{1}\longrightarrow L_{\alpha}\chi)-\Gamma(\Phi_{1}\longrightarrow L_{\alpha}^{c}\chi^{c}) & \equiv & -\epsilon\Gamma_{D} \label{eq:2ndeps} \\
  \Gamma(\overline{\Phi}_{1}\longrightarrow L_{\alpha}^{c}\chi^{c})-\Gamma(\overline{\Phi}_{1}^{c}\longrightarrow L_{\alpha}\chi) & \equiv & \epsilon\Gamma_{D} \label{eq:3rdeps} \\
  \Gamma(\overline{\Phi}_{1}\longrightarrow N_{\alpha}H_{u})-\Gamma(\overline{\Phi}_{1}^{c}\longrightarrow N_{\alpha}^{c}H_{u}^{c}) & \equiv & -\epsilon\Gamma_{D},
  \label{eq:4theps}
\end{eqnarray}
\normalsize where \(\Gamma_{D}\) is the total decay width of any of
the heavy fields in the \(\Phi_{1}\) or \(\overline{\Phi}_{1}\)
supermultiplets, and we have used the superfield notation for
\(\Phi_{1}\), \(N\), etc.\ to indicate that the decay asymmetries
are the same for all supersymmetrizations of the diagrams in
fig.~\ref{fig:DecayDiagrams}.  Explicit calculation of
\(\Gamma_{D}\) and \(\epsilon\) yields
\begin{equation}
  \Gamma_{D}=\frac{1}{16\pi}M_{\Phi_{1}}
  \sum_{\alpha}\left(|\lambda_{1\alpha}|^{2}+|h_{1\alpha}|^{2}\right).
  \label{eq:GammaD}
\end{equation}
and
\begin{equation}
  \epsilon=\frac{\mathrm{Im}(\lambda^{\ast}_{1\alpha}\lambda_{2\alpha}h^{\ast}_{1\beta}h_{2\beta}M_{\Phi_{1}}M^{\ast}_{\Phi_{2}})}{4\pi(|M_{\Phi_{2}}|^{2}-|M_{\Phi_{1}}|^{2})(|\lambda_{1\gamma}|^{2}+|h_{1\gamma}|^{2})},
  \label{eq:epsilon}
\end{equation}
where in both equations, a sum over the repeated indices \(\alpha\),
\(\beta\) and \(\gamma\) is assumed.  Defining
\(\delta\equiv|M_{\Phi_{1}}|/|M_{\Phi_{2}}|\), \(\epsilon\) can be
expressed in the more revealing form
\begin{equation}
  \epsilon=\frac{\mathrm{Im}(\lambda^{\ast}_{1\alpha}\lambda_{2\alpha}h^{\ast}_{1\beta}h_{2\beta}e^{i\psi})}{4\pi(|\lambda_{1\gamma}|^{2}+|h_{1\gamma}|^{2})}\left(\frac{\delta}{1-\delta^{2}}\right),\label{eq:epsilond}
\end{equation}
where \(\psi\) is the relative phase between \(M_{\Phi_{1}}\) and
\(M_{\Phi_{2}}\).  This tells us that for small values of
\(\delta\), the final baryon-to-photon ratio will be approximately
proportional to \(\delta\). Since \(\ell_{\alpha}\) and
\(n_{R\alpha}\) have equal and opposite charges under the global
\(U(1)_{L}\) symmetry, the individual \(L_{\ell}\) and
\(L_{\nu_{R}}\) lepton numbers respectively stored in left-handed
leptons and right-handed neutrinos will likewise be equal and
opposite. Thus no net lepton number is produced by the decays of
\(\Phi\) and \(\overline{\Phi}\); the generation of both
\(L_{\mathit{tot}}\neq0\) and \(B\neq0\) occurs via electroweak
sphaleron processes, which conserve \(B-L_{\ell}\) but violate
\(B+L_{\ell}\). As these are associated with the \(SU(2)\times
U(1)_{Y}\) electroweak anomaly, they will only affect \(L_{\ell}\)
and not \(L_{\nu_{R}}\), and thus create both a nonzero value for
\(B\) and a disparity between the two stores of lepton number (see
fig.~\ref{fig:BLSpaceDiagram1}).

\indent In order for these two stores of lepton number not to be
equilibrated away through the effective Higgs coupling, the
equilibration rate must not become significant compared to the
expansion rate of the universe until after the electroweak phase
transition \(T_{c}\), at which point sphaleron interactions have
effectively turned off.  This equilibration rate may be estimated on
dimensional grounds to be
\begin{equation}
  \Gamma_{\mathit{eq}}\sim
  \frac{|\lambda|^{2}|h|^{2}\langle\chi\rangle^{2}}{M_{\Phi_{1}}^{2}}g^{2}T,
\end{equation}
where \(g\) is an \(\mathcal{O}(1)\) gauge or top Yukawa coupling
and \(T\) is temperature.  The expansion rate may be expressed by
the Hubble parameter \(H=1.66 g_{\ast}^{1/2} T^{2}/M_{P}\), where
\(g_{\ast}\) is the number of interacting degrees of freedom (in the
minimal supersymmetric model (MSSM) during the baryogenesis epoch,
\(g_{\ast}\approx205\)).  Requiring that \(\Gamma_{\mathit{eq}}<H\)
for \(T>T_{c}\) leads to the condition
\begin{equation}
  \frac{|\lambda||h|\langle\chi\rangle}{M_{\Phi_{1}}}\leq10^{-8},
  \label{eq:EquilibrationBound}
\end{equation}
which can be translated into a limit on the neutrino mass using
equation~(\ref{eq:NeutMassMatrixGaugeBasis}): \beq m_\nu\leq 1.74
\sin\beta\ {\rm keV}. \eeq This condition has to be met in any case
because of cosmological constraints on the absolute neutrino mass,
and therefore left-right equilibration would happen well after
sphalerons have shut off, and Dirac neutrinogenesis can then yield
an appropriate value for \(\eta\).

\indent As we mentioned earlier, an important advantage of Dirac
leptogenesis is its versatility: there is nothing in the theory that
fixes any of the parameters \(M_{\Phi_{1}}\), \(M_{\Phi_{2}}\),
\(\lambda_{i\alpha}\), \(h_{i\alpha}\), and \(\langle\chi\rangle\)
to any physical scale.  There are, however numerous constraints on
these parameters.  First, the neutrino masses must be appropriately
small.  Second, we must satisfy the Sakharov criterion that the
abundances of \(\phi_{1}\) and \(\overline{\phi}_{1}\) depart from
their equilibrium values. This occurs when \(\Gamma_{D}\) is slower
than the rate of the expansion of the universe at the temperature
\(T\sim M_{\Phi_{1}}\) when \(\phi_{1}\) can no longer be treated as
effectively massless and its abundance begins to fall off, or in
other words
\begin{equation}
  \frac{\Gamma_{D}}{H(M_{\Phi{1}})}=
  9.97\cdot 10^{-2}\sqrt{g_{\ast}}\frac{M_{P}}{M_{\Phi_{1}}}
  \sum_{\alpha}(|\lambda_{1\alpha}|^{2}+|h_{1\alpha}|^{2})\lesssim
  1.
  \label{eq:GammaOverH}
\end{equation}
This constraint also favors small couplings and large
\(M_{\Phi_{1}}\).  However, in addition to these two requirements,
we must ensure that the present value of \(\eta\) satisfy the bounds
in equation~(\ref{eq:WMAPeta}).

\indent
To derive the precise limits this requirement places on the
model parameters, one must solve the full system of Boltzmann
equations, which we do in section~\ref{sec:BoltzmannEquations}.  For
present purposes, a rough estimate can be made using the ``drift and
decay'' approximation~\cite{Kolb:1990vq}, in which we assume that
the \(\phi\) decays occur well out of equilibrium and that the
effects of inverse
decays and \(2\leftrightarrow2\) 
processes where $\Delta L_{\ell}\neq0$ are negligible. Including
contributions from both scalar and fermion decays, this gives the
result
\begin{equation}
  L_{\ell}=
  \frac{2\epsilon n^{\mathit{init}}_{\phi_{1}}}{s}=
  \frac{90\epsilon}{\pi^{4}g_{\ast}}K_{2}(1)=
  7.32\times10^{-3}\epsilon,
  \label{eq:DriftAndDecay}
\end{equation}
where \(n_{\phi_{1}}^{\mathit{init}}\) is the initial number density
of \(\phi\) and \(K_{2}(x)\) denotes the modified Bessel function of
the second kind, evaluated at x. Since the final baryon-to-entropy
ratio \(B\) (related to \(\eta\) by \(B=\eta/7.04\)) generated by
sphaleron processes will be on the same order ($B\simeq 0.35
L_{\ell}$), this can serve as a rough estimate for \(B\). Thus even
if equation~(\ref{eq:GammaOverH}) is satisfied, the final
baryon-to-entropy ratio of the universe will be proportional to
\(\epsilon\); and from equation~(\ref{eq:epsilon}), we see that for
\(\epsilon\) to be large, either the couplings must be large or the
splitting between \(M_{\Phi_{1}}\) and \(M_{\Phi_{2}}\) must be
small.  Thus there are tensions among the model parameters in Dirac
leptogenesis, but they can be reconciled without too much
difficulty.

\indent The real tensions among \(M_{\Phi_{1}}\), \(M_{\Phi_{2}}\),
\(\lambda_{i\alpha}\), \(h_{i\alpha}\), and \(\langle\chi\rangle\)
are not those inherent in the Dirac neutrinogenesis framework,
however, but those that arise when we demand that the model respect
the full battery of additional constraints from cosmology and from
neutrino physics.  We now turn to address these constraints and
their implications for Dirac neutrinogenesis.

\section{Astrophysical Constraints\label{sec:AstroConstraints}}

\indent

Because the particle spectrum of models with high-scale
supersymmetry breaking may differ significantly from that of models
with weak-scale supersymmetry breaking, it is necessary to
investigate whether any such alterations will affect the
astrophysical constraints (from BBN, inflation, etc.) the theory
must satisfy. In addition, Dirac neutrinogenesis introduces three
additional light, sterile neutrino fields \(\nu_{R\alpha}\), which
could affect BBN. We must make certain that neither of these
considerations presents an insurmountable problem for our model.

\indent We will first address the issue of the additional neutrino
species. It has been shown~\cite{Murayama:2002je} that the presence
of the three additional light neutrino fields \(\nu_{R\alpha}\) at
the time of BBN does not violate the bound~\cite{Olive:1999ij} on
the number of additional light neutrino species \(N_{\nu}\leq 0.3\),
since their contribution to \(N_{\nu}\) is suppressed by a large
entropy factor.  Still, there are other potential cosmological
problems that might arise for split supersymmetry: in particular,
constraints on the reheating temperature \(T_{R}\) associated with
cosmic inflation become constraints on \(M_{\Phi_{1}}\) in thermal
leptogenesis models.  We must ensure that such constraints do not
make Dirac neutrinogenesis unworkable.

\indent The most severe constraints on \(T_{R}\) are rooted in
gravitino cosmology. There are two distinct varieties of gravitino
problem that must be addressed: first, late gravitino decays can
disrupt BBN by releasing energy in the form of photons and other
energetic particles into the system; second, as the gravitino decays
predominately into the LSP and its standard model superpartner, the
potentially large, non-thermal population of stable particles
produced in this manner could overclose the universe---or if the
right amount is produced, could make up the majority of cold dark
matter (CDM).  Both of these issues are contingent on the gravitino
lifetime \(\tau_{3/2}\), which for cases where \(m_{3/2}\gg
m_{LSP}\), is approximated by~\cite{Moroi:1995fs}
\begin{equation}
  \tau_{3/2}=4.0\times 10^{8} \left(\frac{m_{3/2}}{100\mathrm{GeV}}\right)^{-3}\mathrm{s}.
\end{equation}
Careful analysis of the BBN constraints~\cite{Kawasaki:2004qu}
reveals that unless \(m_{3/2}\gtrsim10^{5}\)~GeV, \(T_{R}\) cannot
be greater than around \(10^{8}\)~GeV, and for
\(m_{3/2}\lesssim5\times10^{3}\)~GeV, cannot exceed \(10^{6}\)~GeV.
In models where \(m_{3/2}\) is at the PeV scale, however,
\(\tau_{3/2}\sim 10^{-4}\) s, which implies that gravitinos produced
in the thermal bath decay long before the BBN epoch (at
\(t_{\mathit{universe}}\sim 1\) s), and thus there is no gravitino
problem of the former type for models with \(m_{3/2}\) at or above
these scales. However, since a weakly interacting LSP decouples on a
timescale \(t_{f}\sim 10^{-11}\) s, it will have long frozen out by
the time gravitino decay occurs unless \(m_{3/2}\) is larger than
around \(10^{8}\)~GeV; hence in theories with smaller gravitino
masses (including simple PeV-scale supersymmetry with
anomaly-mediated gaugino masses), LSPs produced by gravitino decay
will be unable to thermalize and the latter type of gravitino issue
cannot be ignored.

\indent In order to avoid any complications from late gravitino
decay, we require not only that the LSP not overclose the universe,
but that its surviving relic density \(\Omega_{\mathit{LSP}}\) must
be less than (or ideally, if the LSP is to constitute the majority
of cold dark matter, equal to) the relic density of CDM as measured
by WMAP~\cite{Spergel:2003cb},
\begin{equation}
  \Omega_{\mathrm{CDM}}h^{2}=0.11\pm0.01 \mbox{ (WMAP 68\% C.L.)}.
  \label{eq:WMAPBounds}
\end{equation}
In general, \(\Omega_{\mathit{LSP}}\) will have both a thermal and a non-thermal component,
so that \(\Omega_{\mathit{LSP}}=\Omega^{\mathit{Th}}_{\mathit{LSP}}+
\Omega^{\mathit{NT}}_{\mathit{LSP}}\).  The thermal component
\(\Omega^{\mathit{Th}}_{\mathit{LSP}}\) may be ascertained by solving the relevant set of
Boltzmann equations for the LSP abundance at freeze-out.  The results, for the case where
the LSP is essentially either a pure Wino or Higgsino, are~\cite{Giudice:2004tc}
\begin{eqnarray}
  \Omega^{\mathit{Th}}_{\mathit{LSP}}h^{2}=0.02\left(\frac{|M_{2}|}{\mathrm{TeV}}\right)
    & \mbox{ for Wino LSP} \\
  \Omega^{\mathit{Th}}_{\mathit{LSP}}h^{2}=0.09\left(\frac{|\mu|}{\mathrm{TeV}}\right)
    & \mbox{ for Higgsino LSP}.
\end{eqnarray}

\indent Now we turn to evaluating
\(\Omega^{\mathit{NT}}_{\mathit{LSP}}\).  We begin by addressing the
regime in which there is no significant reduction in
\(\Omega_{\mathit{LSP}}^{\mathit{NT}}\) from LSP annihilations.
Assuming for the moment that the dominant contribution to the
non-thermal relic abundance comes from late gravitino decays and
that all the LSPs produced from such decays survive until present
day, \(\Omega^{\mathit{NT}}_{\mathit{LSP}}\) is given by
\begin{equation}
  \Omega^{\mathit{NT}}_{\mathit{LSP}}=
  \frac{m_{LSP}\zeta(3)T_{0}^{3}}{\pi^{2}\rho_{\mathit{crit}}}
  Y_{3/2}(T_{3/2}),
  \label{eq:OmegaNTForm}
\end{equation}
where \(\rho_{\mathit{crit}}\) is the critical density of the
universe, \(T_{0}\) is the present temperature of the universe, and
\(Y_{3/2}(T_{3/2})\) is the number of gravitinos per co-moving
volume at the characteristic temperature \(T_{3/2}\) at which the
gravitino decays, which is given by~\cite{Kawasaki:1994af}
\begin{equation}
  Y_{3/2}(T_{3/2})=0.856\times 10^{-11}
  \left(\frac{T_{R}}{10^{10}\mathrm{GeV}}\right)
  \left(1-0.0232\ln\left(\frac{T_{R}}{10^{10}\mathrm{GeV}}\right)\right).
\end{equation}
Substituting this into equation (\ref{eq:OmegaNTForm}) yields
\begin{equation}
  \Omega^{\mathit{NT}}_{\mathit{LSP}}h^{2}=2.96\times 10^{-4}
  \left(\frac{m_{\mathit{LSP}}}{\mathrm{GeV}}\right)
  \left(\frac{T_{R}}{10^{10}\mathrm{GeV}}\right)
  \left(1-0.0232\ln\left(\frac{T_{R}}{10^{10}\mathrm{GeV}}\right)\right).
  \label{eq:OmegaNTNot}
\end{equation}

\begin{figure}[ht!]
  \begin{center}
    \includegraphics{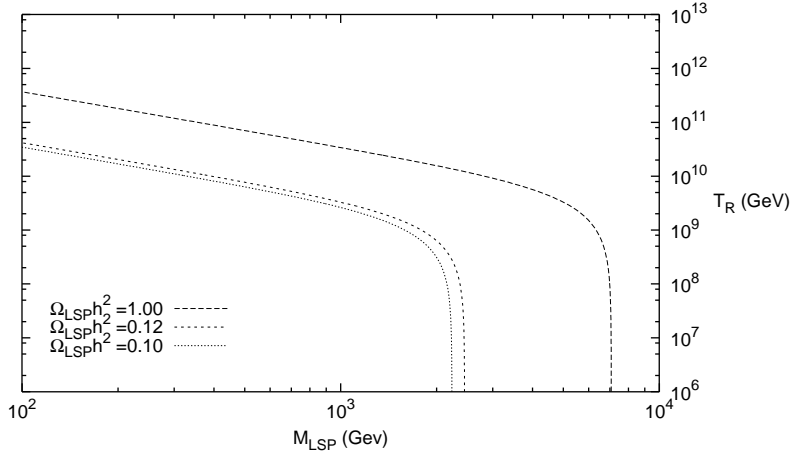}
  \end{center}
  \caption{Contours of \(\Omega_{LSP}h^{2}\) corresponding to the upper and lower bounds from WMAP
(\ref{eq:WMAPBounds}), as well as a simple overclosure bound, as a
function of the LSP mass \(m_{\mathit{LSP}}\) and the reheating
temperature \(T_{R}\) associated with cosmic inflation.  The
requirement that \(\Omega_{\mathit{LSP}}\) not overclose the
universe severely constrains \(T_{R}\), and hence the temperature
scale of thermal leptogenesis, for a theory like PeV-scale
loop-split supersymmetry, in which the LSP is particularly
heavy.\label{fig:OmegaTot}}
\end{figure}

\indent In fig.~\ref{fig:OmegaTot}, we plot the contours
corresponding to the WMAP upper and lower bounds from equation
(\ref{eq:WMAPBounds}) on the total LSP relic abundance, as well as
the simple overclosure bound \(\Omega_{\mathit{LSP}}h^{2}=1\),
taking into account both thermal and non-thermal contributions, as a
function of \(m_{\mathit{LSP}}\) and \(T_{R}\).  The region above
and to the right of the WMAP upper bound is excluded.  In order for
the LSP to make up a significant fraction of the dark matter,
\(\Omega_{\mathit{LSP}}\) should fall within the narrow strip
between the upper and lower bounds on \(\Omega_{\mathit{CDM}}\)
given in~(\ref{eq:WMAPBounds}), but if some other particle makes up
the majority of CDM, then the entire region below and to the left of
the WMAP upper bound is phenomenologically allowed.  In order to
achieve an appropriate LSP relic density by thermal means alone, it
can be seen that \(T_{R}\) must be quite low--around \(10^{9}\) GeV;
if the majority of CDM is generated non-thermally, \(T_{R}\) may be
raised a bit, but is still constrained to be below \(\sim 5\times
10^{10}\)~GeV.  As we shall see in
section~\ref{sec:BoltzmannEquations}, \(T_{R}\sim10^{9}\) GeV turns
out to be problematic for Dirac neutrinogenesis (in terms of the
final baryon-to-photon ratio generated), largely due to the out-of
equilibrium condition in~(\ref{eq:GammaOverH}), this implies that if
we want to raise \(T_{R}\) above \(10^{9}\)~GeV and still have the
LSP relic density dominate \(\Omega_{\mathit{CDM}}\), the dark
matter must be essentially non-thermal in origin.

\indent We now turn to address the regime where LSP annihilations do
play a role in reducing \(\Omega_{\mathit{LSP}}^{\mathit{NT}}\), and
thus the upper bound on \(T_{R}\) may be raised.  This effect
becomes important when \(m_{3/2}\gg m_{\mathit{LSP}}\).  When it is
taken into account~\cite{Moroi:1999zb}, the non-thermal LSP relic
density is modified to
\begin{equation}
  \Omega^{\mathit{NT}}_{\mathit{LSP}}=\min
    \left(\Omega^{\mathit{NT} (0)}_{\mathit{LSP}},\Omega^{\mathit{NT} \mathit{(ann)}}_{\mathit{LSP}}\right),
\end{equation}
where \(\Omega^{\mathit{NT} (0)}_{\mathit{LSP}}\) is the relic density given in equation
(\ref{eq:OmegaNTNot}), and \(\Omega^{\mathit{NT} \mathit{(ann)}}_{\mathit{LSP}}\) is the
relic density obtained by solving the full system of Boltzmann equations for the LSP.  For
a Wino LSP, \(\Omega^{\mathit{NT} \mathit{(ann)}}_{\mathit{LSP}}\) is given by
\begin{eqnarray}
  \Omega^{\mathit{NT} \mathit{(ann)}}_{\mathit{LSP}} & = &
  2.41\times10^{-2}
  \frac{(2-x_{W})^{2}}{(1+x_{W})^{3/2}}
  \left(\frac{m_{\mathit{LSP}}}{100\mathrm{GeV}}\right)^{3}
  \left(\frac{m_{3/2}}{100\mathrm{TeV}}\right)^{-3/2} \nonumber \\ & \times &
  \left(1-\left(\frac{m_{\mathit{LSP}}}{m_{3/2}}\right)\right)^{3}
  \left(1+\frac{1}{3}\left(\frac{m_{\mathit{LSP}}}{m_{3/2}}\right)\right),
  \label{eq:OmegaNTAnn}
\end{eqnarray}
where \(x_{W}\equiv m_{W}/m_{\mathit{LSP}}\); for a Higgsino LSP,
which annihilates far less efficiently, \(\Omega^{\mathit{NT}
(\mathit{ann})}_{\mathit{LSP}}\) will be even higher.

\begin{figure}[ht!]
\hspace{1.5cm}
  \includegraphics{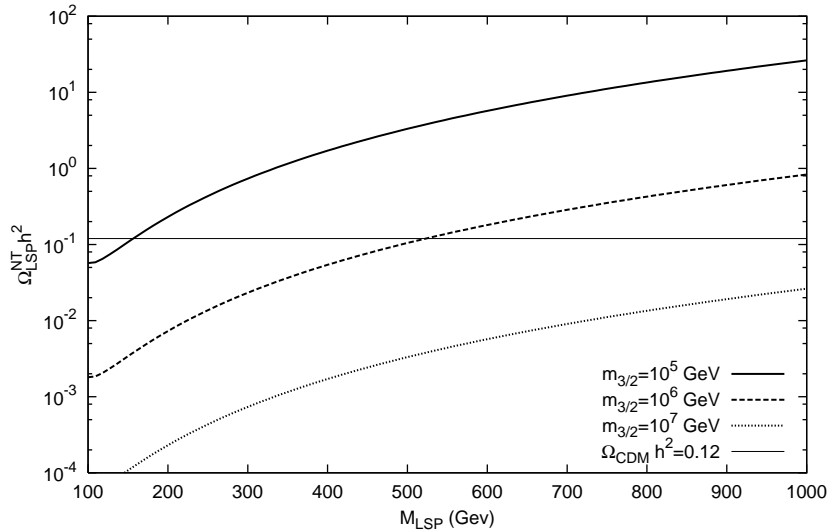}
  \caption{The variation of \(\Omega^{\mathit{NT} \mathit{(ann)}}_{\mathit{LSP}}\) (the LSP
    relic density function when LSP annihilations are taken into account) with LSP mass for
    several different values of \(m_{3/2}\).  When \(m_{3/2}=10^{6}\)~GeV, only for
    \(m_{LSP}\leq500\) GeV will \(\Omega^{\mathit{NT} \mathit{(ann)}}_{\mathit{LSP}}\) not
    generate an LSP relic density that exceeds the WMAP upper bound on the CDM relic density,
    and thus only for such an LSP mass can the reheating temperature \(T_{R}\) be increased beyond
    the naive upper bound from fig.~\ref{fig:OmegaTot}.
    For larger gravitino masses \(m_{3/2}\geq10^{7}\)~GeV, annihilations effectively reduce
    \(\Omega_{\mathit{CDM}}^{\mathit{NT}}\) and allow for \(T_{R}\) to be increased above the
    naive upper bound.\label{fig:OmegaAnn}}
\end{figure}

\indent In fig.~\ref{fig:OmegaAnn}, we show the relationship between
\(\Omega^{\mathit{NT} (\mathit{ann})}_{\mathit{LSP}}\) and
\(m_{\mathit{LSP}}\) for several values of \(m_{3/2}\).  To be of
any benefit in reducing \(\Omega^{\mathit{NT}}_{\mathit{LSP}}\),
\(\Omega^{\mathit{NT} \mathit{(ann)}}_{\mathit{LSP}}\) itself must
not exceed the WMAP upper bound, which is included for reference.
From this plot it is evident that annihilations are only effective
in reducing the LSP relic abundance below this bound when
\(m_{3/2}\) is much larger than \(m_{\mathit{LSP}}\).  This is a
problem for loop-split supersymmetry with \(m_{3/2}\) around the PeV
scale, where the ratio of \(m_{LSP}\) to \(m_{3/2}\) is explicitly
determined by the anomaly-mediation relation~(\ref{eq:AMSBMass}). It
is not, however, a problem for more general split SUSY scenarios, in
which the splitting between \(m_{LSP}\) and \(m_{3/2}\) can be much
larger~\cite{Kawasaki:1994af,Ibe:2004tg}. Furthermore, when
\(m_{3/2}\) is increased beyond around \(10^{8}\), \(\tau_{3/2}\)
becomes short enough that gravitino decay occurs before LSP
freeze-out, and \(\Omega_{\mathit{LSP}}^{\mathit{NT}}\) drops to
zero regardless of what the ratio of \(m_{\mathit{LSP}}\) to
\(m_{3/2}\) is.

\indent Since thermal leptogenesis requires that a thermal
population of \(\Phi_{1}\) and \(\overline{\Phi}_{1}\) be generated
during reheating after inflation, we require that
\(M_{\Phi_{1}}\lesssim T_{R}\); hence an upper bound on \(T_{R}\)
translates into an upper bound on \(M_{\Phi_{1}}\).  This means that
there is substantial model tension when the \(T_{R}\) is constrained
to be at or below \(\sim5\times10^{10}\)~GeV. This is not in any way
peculiar to Dirac leptogenesis either: in a Majorana leptogenesis
model, the constraints still apply with the lightest right-handed
neutrino mass \(M_{\nu_{R}}\) in place of \(M_{\Phi_{1}}\).  Here,
the out-of-equilibrium condition~(\ref{eq:GammaOverH}) demands that
\(\lambda_{1\alpha}\) and \(h_{1\alpha}\) be at most
\(\mathcal{O}(10^{-4})\) for \(M_{\Phi_{1}}\sim10^{10}\) GeV; on the
other hand, (\ref{eq:epsilon}) and~(\ref{eq:DriftAndDecay}) imply
that unless there is a large hierarchy among the Yukawa couplings to
different sets of \(\Phi\) and \(\overline{\Phi}\) (so that
\(\lambda_{1\alpha}\ll\lambda_{2\alpha}\) for all $\alpha$), or
unless \(M_{\Phi_{1}}\) and \(M_{\Phi_{2}}\) are essentially
degenerate, making the \(\lambda_{i\alpha}\) and \(h_{i\alpha}\) any
smaller than \(\mathcal{O}(10^{-4})\) will yield insufficient baryon
number---and we haven't yet taken into account the effects of
\(2\leftrightarrow2\) processes and inverse decays. In short,
leptogenesis becomes difficult when there are upper bounds on
\(T_{R}\), which occurs when \(m_{3/2}\lesssim10^8\)~GeV and LSP
annihilations are ineffective.

\indent There are two methods for increasing the baryon density in
models where \(M_{\Phi_{1}}\) is tightly constrained by bounds on
the reheating temperature, as indicated in
equation~(\ref{eq:epsilon}). One possibility, to which we have
already alluded, is somehow to arrange a large hierarchy among the
Yukawa couplings to different sets of \(\Phi\) and
\(\overline{\Phi}\) (i.e. to adjust the matrixes \(\lambda\) and
\(h\)). Since the couplings to \(\Phi_{2}\) only influence the
physics via their \(CP\)-violating phases in \(\epsilon\)
(\ref{eq:epsilon}) and via their contribution to the mass matrix
(\ref{eq:NeutMassMatrixGaugeBasis}), increasing them will cause no
phenomenological problems as long as a realistic neutrino mass
spectrum is obtained.  The second possibility is to make
\(M_{\Phi_{1}}\) and \(M_{\Phi_{2}}\) very close together in order
to achieve a resonance condition in \(\epsilon\) (i.e. to adjust
\(\delta\), the ratio of these two quantities). Since the
perturbation theory we have used in calculating equation
(\ref{eq:epsilon}) is good as long as the separation of
\(M_{\Phi_{1}}\) and \(M_{\Phi_{2}}\) is substantially greater than
the value of the off-diagonal elements \(M_{ij}\) in the
\(\Phi-\overline{\Phi}\) mass mixing matrix induced at the one-loop
level:
\begin{equation}
  M_{ij}\sim g^{\ast} g' (M_{\Phi_{1}}+M_{\Phi_{1}})I_{ij},
\end{equation}
where \(g_{i}\) and \(g'_{j}\) represent the appropriate \(\lambda\)
and \(h\), summed over the fermion family index, and \(I_{ij}\) is a
numerical factor on the order of \(1/16\pi^{2}\) from the loop
integral. Thus for small \(\lambda\) and \(h\), \(\delta\) can be
set very close to one and \(M_{\phi_{1}}\) and \(M_{\phi_{2}}\) may
be very nearly degenerate.  This allows for the possibility of
resonant leptogenesis, as has been done in the Majorana leptogenesis
case, which can be invoked to generate a large baryon number in
cases where the bounds on \(T_{R}\) are more severe.  As we shall
see in the next section, scenarios with small \(\delta\) can
naturally produce a realistic neutrino spectrum; thus we will
restrict our attention to the case where \(\delta\) is small.

\indent While the problems that can arise for small gravitino masses
have now been thoroughly addressed, the caveats associated with
extremely large \(m_{3/2}\) should also be mentioned. As has been
shown in~\cite{Arvanitaki:2005fa}, split supersymmetry models with a
large hierarchy between the gravitino and gaugino masses can suffer
from phenomenological problems associated with the overproduction of
gluinos.  While a precise ceiling for \(m_{3/2}\) must wait until
gluino decay is better understood, it is known that this ceiling
falls somewhere in the \(m_{3/2}\simeq10^{10} - 10^{12}\) GeV range.
There are also caveats associated with the gravitino-producing
decays of scalar sparticles in models where one or more scalars has
a mass larger than \(m_{3/2}\)~\cite{Allahverdi:2005rh}.

\indent The message here is that cosmological considerations can
make thermal Dirac neutrinogenesis substantially more difficult when
\(m_{3/2}\lesssim 10^{8}\)~GeV, as substantial tensions arise among
the out-of-equilibrium decay criterion, overclosure bounds related
to the reheating temperature \(T_{R}\), the equation that determines
the decay asymmetry \(\epsilon\), etc.  For light gravitinos
(\(m_{3/2}\lesssim10^{5}\)~GeV), BBN limits severely constrain
\(T_{R}\), and hence \(M_{\Phi_{1}}\).  For slightly heavier
gravitinos (\(10^{5}\)~GeV~\(\lesssim m_{3/2}\lesssim10^{8}\)~GeV),
there are still constraints on \(T_{R}\) from nonthermal decays
which are only alleviated (via LSP annihilations) when
\(m_{\mathit{LSP}}\ll m_{3/2}\).  Since this is not the case in
loop-split supersymmetry, where the ratio of \(m_{\mathit{LSP}}\) to
\(m_{3/2}\) is specified, making Dirac neutrinogenesis succeed is
quite difficult in this framework. Still, if we allow for
non-thermal generation of \(\Omega_{\mathit{CDM}}\) (or for the
possibility that the LSP is not a significant contributor to the
dark matter density), Dirac neutrinogenesis may still be viable in
loop-split SUSY (while there are apparent tensions in this case,
they an be resolved by invoking some mechanism, such as resonant
leptogenesis, for amplifying the lepton number produced).  In more
general split supersymmetry scenarios, however, \(T_{R}\) and
\(M_{\Phi_{1}}\) may be increased to \(10^{12}\) GeV or more,
leptogenesis should proceed without too much trouble, and the
possibility for thermal dark matter still exists. Whatever the case,
it will be necessary to solve the full system of Boltzmann equations
to ascertain whether or not there is any region of parameter space
in which Dirac neutrinogenesis can be successful, which we do in
section~\ref{sec:BoltzmannEquations}.

\section{Neutrino Physics\label{sec:NeutSpec}}

\indent

In addition to respecting constraints arising from cosmological
considerations, in order to be phenomenologically viable, a given
Dirac neutrinogenesis model must yield a neutrino spectrum that
accords with current experimental constraints.  The most stringent
such constraints come from solar and atmospheric neutrino
oscillation experiments, and place limits both on the mass
splittings
\begin{equation}
  \Delta m_{ab}^{2}\equiv m_{\nu_{a}}^{2}-m_{\nu_{b}}^{2},
\end{equation}
where the indices \(a\) and \(b\) label the different neutrino mass eigenstates, and on the
mixing angles \(\theta_{ab}\) between these eigenstates.  The primary connection between the
latter and observable physics occurs through the leptonic mixing matrix
\begin{equation}
  U_{\mathit{MNS}}=U^{(\nu)}_{}U^{(e)}_{},
  \label{eq:UMNSDef}
\end{equation}
where \(U^{(\nu)}_{}\) is the neutrino mixing matrix and \(U^{(e)}_{}\) is the charged
lepton mixing matrix. In the basis where the charged lepton mass matrix is diagonal,
\(U_{\mathit{MNS}}\) is the neutrino mixing matrix and may be expressed in terms of the neutrino mixing angles
\(\theta_{ab}\) as
\begin{equation}
  U_{\mathit{MNS}}=
  \left(\begin{array}{ccc} c_{12}c_{13} & s_{12}c_{13} & s_{13}e^{i\delta_{\mathit{CP}}} \\ -s_{12}c_{23}-c_{12}s_{23}s_{13}e^{i\delta_{\mathit{CP}}} & c_{12}c_{23}-s_{12}s_{23}s_{13}e^{-i\delta_{\mathit{CP}}} & s_{23}c_{13} \\ s_{12}s_{23}-c_{12}c_{23}s_{13}e^{-i\delta_{\mathit{CP}}} & -c_{12}s_{23}-s_{12}c_{23}s_{13}e^{-i\delta_{\mathit{CP}}} & c_{23}c_{13} \end{array}\right),
  \label{eq:UMNS}
\end{equation}
where \(c_{ab}=\cos\theta_{ab}\), \(s_{ab}=\sin\theta_{ab}\), and
\(\delta_{\mathit{CP}}\) is a \(CP\)-violating phase.  The present
limits\footnote{We do not take into account the LSND result which
would require an extra neutrino mass eigenstate. In the case that
forthcoming data from experiments such as MiniBooNE corroborate the
LSND signal, it will be necessary to extend the neutrino content of
our model.} on the \(\Delta m_{ab}^{2}\) and \(\theta_{ab}\)
are~\cite{Hagedorn:2005kz}
\begin{eqnarray}
  \begin{array}{ccc}
    \sin^{2}\theta_{12}=0.30^{+0.04}_{-0.05}, & \sin^{2}\theta_{23}=0.50^{+0.14}_{-0.12}, &
    \sin^{2}\theta_{13}\leq0.031,
  \end{array} \nonumber \\
  \begin{array}{cc}
    \Delta m_{21}^{2}=\left(7.9^{+0.6}_{-0.6}\right)\times 10^{-5} \mathrm{eV}^{2}, &
    |\Delta m_{31}^{2}|=\left(2.2^{+0.7}_{-0.5}\right)\times 10^{-3} \mathrm{eV}^{2}.
  \end{array}
  \label{eq:NeutLimits}
\end{eqnarray}
The smaller of the two mass splittings, \(\Delta m_{21}^{2}\), is to
be identified with the \(\Delta m_{\odot}^{2}\) obtained from solar
neutrino data (the MSW-LMA solution); the larger, \(\Delta
m_{21}^{2}\), with the \(\Delta m_{A}^{2}\) from atmospheric
neutrino data. {The constraints in~(\ref{eq:NeutLimits}) imply a
\(U_{\mathit{MNS}}\) matrix with bounds
\begin{equation}
  |U_{\mathit{MNS}}|=\left(\begin{array}{ccc} .79-.86 & .49-.58 & 0-.18 \\ .30-.58 & .40-.68 & .61-.80 \\ .19-.46 & .50-.77 & .59-.79 \end{array}\right),\label{eq:MNSdata}
\end{equation}
but they leave the sign of the largest mass squared difference
$\Delta m_{31}^{2}$ undefined. This implies that the physical
neutrino $\nu_3$ can be either the heaviest of the three mass
eigenstates, i.e. $m_1<m_2\ll m_3$ (normal hierarchy) or the
lightest, i.e. $m_3\ll m_1<m_2$ (inverted hierarchy).

The unitary matrix $U_{\mathit{MNS}}$, in the charged lepton basis,
is responsible for diagonalizing the squared neutrino mass matrix ,
i.e. $$\left(m_\nu^2\right)^{diag}=\ U_{\mathit{MNS}}^\dagger\ m_\nu
m_\nu^\dagger\ U_{\mathit{MNS}}.$$ We can therefore estimate
\cite{Hagedorn:2005kz} the generic form of the neutrino mass matrix
squared, since it has to be diagonalized by the $U_{\mathit{MNS}}$
matrix.

We should have for a Normal Hierarchy
\begin{equation}
  (m^{2}_{\nu})^{Norm}\sim \Delta m_{31}^{2}
  \left(\begin{array}{ccc}\xi&\xi&\xi\\ .&1&1\\ .&.&1\end{array}\right),
    \label{eq:NeutMassNormalGBForm}
\end{equation}
and for an Inverted Hierarchy
\begin{equation}
  (m^{2}_{\nu})^{Inv}\sim \Delta m_{31}^{2}
  \left(\begin{array}{ccc}1&\xi&\xi\\ .&1&1\\ .&.&1\end{array}\right),
    \label{eq:NeutMassInvertedGBForm}
\end{equation}
where the $\xi$ are small entries (not necessarily equal) compared
to the $\mathcal{O}(1)$ entries. In both cases, the order of
magnitude of $\xi$ is constrained by the ratio \(\rho_{32}\equiv
\Delta m_{31}^{2}/\Delta m_{21}^{2}\), which, according
to~(\ref{eq:NeutLimits}), must respect the bounds \beq
  20.0 < \rho_{23} < 39.7\ .
\eeq
We are interested in finding a simple connection between the heavy
$\Phi$ fields and the neutrino sector, which does generically
reproduce the observed neutrino spectrum and is also compatible
with successful baryon number generation.


\indent
Let us first consider
the spectrum of a model with only one set of \(\phi\) and \(\overline{\phi}\), or
alternatively, a
theory in which \(M_{\Phi_{1}}\ll M_{\Phi_{i}}\) for all \(i>1\).  In such a case, the mass
matrix is proportional to an outer product of the family-space vectors \(\lambda_{1\alpha}\)
and thus its eigenvalues are
\begin{eqnarray}
  m_{\nu_{1}}=0 & m_{\nu_{2}}=0 &
  m_{\nu_{3}}=\sum_{\alpha}^{3}\lambda_{1\alpha}h_{1\alpha}.
\end{eqnarray}
For each additional set of \(\phi\) and \(\overline{\phi}\) with a
mass similar to \(M_{\phi_{1}}\), an additional neutrino acquires a
nonzero mass.  Thus in ``short-suited'' models in which there are
effectively only two sets of \(\phi\) and \(\overline{\phi}\)
involved in determining the neutrino spectrum, one neutrino mass
eigenstate is massless, and the other two neutrino masses are given
by \(m^{2}_{\nu_{2}}\approx\Delta m^{2}_{21}\) and
\(m^{2}_{\nu_{3}}\approx\Delta m^{2}_{31}\).  Furthermore,
leptogenesis in such models is well-approximated by the toy model
considered in section~\ref{sec:DiracLep}, in which there were only
two sets of \(\Phi\) and \(\overline{\Phi}\).

{
\indent
From~(\ref{eq:NeutMassMatrixGaugeBasis}), the neutrino mass-squared
matrix
is given by

\begin{equation}
  |m_{\nu}|^{2}_{\alpha\beta}= \Big(v \langle\chi\rangle \sin\beta\Big)^2\
  \sum_{i,j=1}^{2\ (or\ 3)} \sum_{\gamma=1}^3\lambda^{\ast}_{i\gamma}\lambda_{j\gamma}
  h^{\ast}_{i\alpha } h_{j\beta} \frac{1}{M^\ast_{\Phi_i}M_{\Phi_j}}.
  \label{eq:NeutMassSpelledOut}
\end{equation}
As long as \(\lambda\) and \(h\) are completely generic and there
are at least three sets of \(\phi\) and \(\overline{\phi}\), it is
apparent that a matrix of this form can yield an arbitrary neutrino
mass spectrum.  On the one hand this is good, for it means that the
theory is perfectly viable; on the other hand, this arbitrariness
comes at the price of introducing many additional free parameters,
whose relative values must be determined by some additional
underlying physics.

Guided by the idea of linking the neutrino data to the $\Phi$ fields
sector, we find two simple conditions that generically reproduce the
form of the {\it normal hierarchy} neutrino mass squared
matrix~(\ref{eq:NeutMassNormalGBForm}).

\bit
\item {\it Normal Hierarchical ansatz A}

\indent
A quick inspection of (\ref{eq:NeutMassSpelledOut}) shows
that if the spectrum of heavy fields is hierarchical, i.e. if
$M_{\Phi_1}\ll M_{\Phi_2}(\ll M_{\phi_3})$\footnote{Of course $M_{\Phi_2}$ cannot be
  too large since then, as remarked earlier, we effectively
  generate the mass of just one physical neutrino leaving the other two massless.}
then we can very simply write the neutrino mass squared matrix to
zero order in $\delta=M_{\Phi_1}/M_{\Phi_2}$ as \bea
|m_\nu|^2=\Big(v \langle\chi\rangle \sin\beta\Big)^2 {\Lambda_1\over
M^2_{\Phi_1}}\
\left(\begin{array}{ccc}|h_{11}|^2&h^{\ast}_{11}h_{12}&h_{11}^{\ast}h_{13}\\
.&|h_{12}|^2&h_{12}^{\ast}h_{13}\\ .&.&|h_{13}|^2\end{array}\right)
+ {\cal O}(\delta),\label{eq:hierarA} \eea where
$\Lambda_1=\sum_\gamma |\lambda_{1\gamma }|^2 $.

Another hierarchical assumption, namely the requirement that the
off-diagonal elements of the coupling matrix $h$ be much greater than the
diagonal elements ($h_{12}\sim h_{13}\equiv \tilde{h}\gg h_{11}$),
allows us to rewrite the previous equation as
\bea
|m_\nu|^2=\left(v  \langle\chi\rangle \sin\beta\right)^2 {|\tilde{h}|^2 \Lambda_1\over M^2_{\Phi_1}}\
\left(\begin{array}{ccc}\varepsilon^2&\varepsilon&\varepsilon\\ .&1&1\\ .&.&1
\end{array}\right)
+ {\cal O}(\delta), \eea where $\varepsilon=h_{11}/\tilde{h}$.

This reproduces the structure required for a {\it normal hierarchy}
neutrino sector.
It is interesting that the two simple hierarchical requirements, $M_{\Phi_1}\ll M_{\Phi_{2,3}}$
and $h_{11}\ll h_{12}\sim h_{13}$ (with $h_{11}\neq 0$), generically
give rise to the correct neutrino phenomenology.

\item {\it Normal Hierarchical ansatz B}

\indent The previous ansatz suggests that we look for some setup
that would ensure that the diagonal element $h_{11}$ is small
compared to the off diagonal terms $h_{12}$ and $h_{13}$. An
antisymmetric condition on the matrix $h$ can do the job, but from
eq.~(\ref{eq:hierarA}) it becomes clear that we need to include a
higher order term in the small parameter $\delta$.

When we assume that all the diagonal elements of \(h\) are zero (or
extremely small), and the rest of elements in $h$ or $\lambda$ are
of order $1$, we obtain the structure \beq (m^{2}_{\nu})\propto
  \left(\begin{array}{ccc}\delta^2&\delta&\delta\\ .&1&1\\ .&.&1\end{array}\right).
\label{eq:DeltasInFirstRow}\eeq Again we find the generic structure
required for a normal hierarchy, but now the value of $\delta$ fixes
the ratio $\rho_{23}$. \eit

Clearly, either ansatz A or B (or a combination of both) can explain
the neutrino data and large mixing angles. Ideally, we would like to
have a theoretical framework that would provide both the form of the
superpotential required for Dirac neutrinogenesis
(\ref{eq:DiracLepSuperpotential}) and the necessary flavor
structure, like the ones presented in the previous examples.  A
careful construction of such a model is out of the scope of this
paper, but we nevertheless would like to point at least in one
interesting direction in which simple assumptions enable us to
reproduce the needed conditions, while at the same time reducing the
number of new free parameters.

In many grand unified theories and models with non-Abelian flavor
symmetries, the Yukawa matrices (and in general the flavor
interactions) can be symmetric, antisymmetric, or both (see for
example~\cite{Davidson:1980sx,King:2004tx,Froggatt:1978nt} and
references therein).  Operating in this paradigm, let us assume that
the SM left-handed leptons have the same flavor charge as the heavy
fields $\bar{\Phi}$, and the SM right-handed charged leptons have
same flavor charge as the fields $\Phi$. Upon breaking of the flavor
symmetry, let us assume that the charged lepton Yukawa matrix is
symmetric due to a symmetric flavon VEV configuration $\langle
S_{\alpha\beta}\rangle$. The corresponding effective superpotential
is \beq W\supset Y_l\ {\langle S_{\alpha \beta}\rangle \over M_F}
H_d L_\alpha e_\beta + Y_{\Phi}\ \langle S_{\alpha \beta}\rangle
\bar{\Phi}_\alpha\Phi_\beta {\rm +h.c.} \eeq where $Y_l$ and
$Y_{\Phi}$ are dimensionless couplings, and $\alpha$ and $\beta$ are
flavor indices. The charged lepton Yukawa matrix
$(y_{{}_l})_{\alpha\beta}$ and the mass matrix
$(M_{\Phi})_{\alpha\beta}$ of the heavy fields $\Phi$ would be both
symmetric and proportional \beq (M_{\Phi})_{\alpha\beta} = M_F
{Y_{\Phi}\over Y_l}\ (y_{{}_l})_{\alpha\beta}\label{eq:MandYProp}
\eeq This specific structure predicts exactly the mass spectrum for
the $\phi$ fields in terms of the flavor scale $M_F$, which is at
the origin of the intermediate scale required for successful thermal
Leptogenesis. \bea M_{\Phi_1}=m_e {M_F\over v}, \hspace{.5cm}
M_{\Phi_2}=m_\mu {M_F\over v} \hspace{.2cm}{\rm and}\hspace{.2cm}
M_{\Phi_3}=m_\tau {M_F\over v} \hspace{.5cm} \eea If the flavor
scale is of the order of some GUT scale $M_F\sim 10^{16}$~GeV, we
would then expect $M_{\Phi_1}\sim 10^{11}$~GeV, with
$\delta=M_{\Phi_1}/M_{\Phi_2}=m_e/m_\mu\sim 5\times 10^{-3}$ being a
small parameter. If one worries about the dependance on the Flavor
symmetry scale it is also possible to link the intermediate scale to
the large SUSY breaking scale $M_{\mathit{SUSY}}$. We can imagine
that if all fields in the theory are charged, the spurion field $X$
of supersymmetry breaking might as well be charged, and perhaps it
is charged under the same symmetry $U(1)_N$ responsible for
obtaining the superpotential (\ref{eq:DiracLepSuperpotential}).
Before any kind of breaking we can imagine the superpotential as
\beq W\supset {S_{\alpha \beta}\over M_F}\left(Y_l\ H_d L_\alpha
e_\beta + Y_{\Phi}\ X\ \bar{\Phi}_\alpha\Phi_\beta\right) {\rm
+h.c.} \eeq where again $Y_l$ and $Y_{\Phi}$ are dimensionless
couplings. Supersymmetry breaking effects can provide the field $X$
with a VEV $\langle A_{X}\rangle\sim \left(m_{3/2}
M_{Pl}\right)^{1/2}$, and upon flavor symmetry breaking we could get
the effective superpotential \beq W_{\mathit{eff}}\supset
(y_{{}_l})_{\alpha\beta}\ \left(H_d L_\alpha e_\beta +\sqrt{m_{3/2}
M_{Pl}}\ \bar{\Phi}_\alpha\Phi_\beta\right) {\rm +h.c.} \eeq where
we have assumed that the original constants $Y_l\sim Y_\Phi$. In
this situation, for example with $m_{3/2}\sim 10^{10}$ GeV, the mass
of the lightest $\Phi$ field would be $M_{\Phi_1}\sim10^{11}$ GeV.

Now, let us also assume that the flavon VEV $ \langle
S_{\alpha\beta}\rangle$ is symmetric and that the coupling
$\bar{\Phi} L \chi$ becomes antisymmetric upon flavor breaking, i.e.
the superpotential can be written as \beq W\supset
(y_{{}_l})^{sym}_{\alpha\beta} H_d L_\alpha e_\beta\ +
(M_\Phi)^{sym}_{\alpha \beta} \bar{\Phi}_\alpha\Phi_\beta\
+\lambda_{\alpha\beta}N_\alpha \Phi_\beta H_u +
h_{\alpha\beta}^{antisym} \bar{\Phi}_\alpha L_\beta \chi \eeq where
we have $\lambda_{\alpha\beta}\equiv \langle
A^N_{\alpha\beta}\rangle$ and $h_{\alpha\beta}^{anti}\equiv \langle
A^\chi_{\alpha\beta}\rangle$, with the flavon $A^N$ acquiring an
antisymmetric VEV configuration and with the VEV configuration of
$A^\chi$ being arbitrary in flavor space (it is not important to our
purposes its flavor structure although probably in a specific model
of flavor it would end up being some linear combination of symmetric
and antisymmetric VEV configurations).
Let us see what the implications of these ingredients are for our
model.

In general, the charged lepton Yukawa matrix
\((y_{{}_l})_{\alpha\beta}\) can be diagonalized by a biunitary
transformation of the type
\begin{equation}
  U^{\dagger}_{}y_{{}_l} V_{}=y_{{}_l}^{\mathit{(diag)}},
\end{equation}
but when \(y_{l}\) is symmetric, this biunitary transformation takes the simpler form
\begin{equation}
    U^{T}_{}y_{{}_l} U_{}=y_{{}_l}^{\mathit{(diag)}}.
\end{equation}
If \((M_{\Phi})\propto  (y_{{}_l})\), as in the setup described
above (see equation~(\ref{eq:MandYProp})), then \((M_{\Phi})\) will
be diagonalized by the same transformation. When the mass matrices
for the charged leptons and the \(\phi-\overline{\phi}\) system are
simultaneously diagonalized (that is, when we go to the charged
lepton basis), \(\lambda\) and \(h\) transform as
\begin{eqnarray}
  \lambda'_{}=U^{T}_{}\lambda_{}U_{} &\hspace{.5cm} &
  {h^{'\ }}^{anti}_{}=U^{T}_{}h^{anti}_{}U_{}.
\end{eqnarray}
Transformations of this type preserve the antisymmetry of \(h\) (and
$\lambda$ if also antisymmetric); thus in the charged lepton basis
the matrix $h$ remains antisymmetric and $(M_{\Phi})$ is real and
diagonal; this reproduces the {\it Hierarchical ansatz B} described
earlier, in which the diagonal elements of $h$ are zero and it is
the smallness of $\delta$ the responsible for the large mixing
angles observed in the lepton mixing matrix $U_{\mathit{MNS}}$.

\indent In the setup discussed above, \((M_{\Phi})\) and $y_{l}$
could be simultaneously diagonalized because \((M_{\Phi})\propto
y_{l}\), which fixes
\(\delta=M_{\Phi_{1}}/M_{\Phi_{2}}=m_{e}/m_{\mu}\).  Such
proportionality is not requires, however: any matrix of the form

\begin{equation}
  (M_{\Phi})=Ay_{l}+BI_{3\times3},
\end{equation}
where $A$ and $B$ are arbitrary constants and \(I_{3\times3}\) it
the \(3\times3\) identity matrix, can be diagonalized along with
$y_{l}$.  Thus when \(B\neq0\), \(\delta\) can in principle take any
value (as long as that value is consistent with the observational
bounds on the neutrino spectrum).

\bit
\item {\it Constrained Hierarchical Dirac Leptogenesis}

\indent Motivated by the preceding remarks, we define Constrained
Hierarchical Dirac Leptogenesis (CHDL) as the phenomenological setup
in which \((M_{\Phi})\) is real and diagonal, $\lambda$ and $h$ are
antisymmetric, and the smallness of
\(\delta=M_{\Phi_{1}}/M_{\Phi_{2}}\) explains the large mixing
angles in the neutrino sector.  We may parameterize \(\lambda\) and
\(h\) as

\begin{eqnarray}
  \lambda=f
  \left(\begin{array}{ccc} 0 & 1 & a_{2}\\-1 & 0 & a_{3} \\ -a_{2} & -a_{3} & 0\end{array}\right) &
  \hspace{.5cm} &
  h=f
  \left(\begin{array}{ccc} 0 & b_{1} & b_{2}\\ -b_{1} & 0 & b_{3} \\ -b_{2} & -b_{3} & 0\end{array}\right),
  \label{eq:YukawaParametrization}
\end{eqnarray}
which is convenient when the \(a\) and \(b\) are all roughly
\(\mathcal{O}(1)\).  Since the assumption of a hierarchy among the
\(M_{\Phi_{i}}\) leads to the neutrino mass-squared matrix of the
form~(\ref{eq:hierarA}), we expect that \(a_{3}\) and \(b_{3}\), the
effects of which show up only at the \(\mathcal{O}(\delta)\) level,
will be less tightly constrained than the rest of the \(a_{i}\) and
\(b_{i}\), which contribute to the leading term.  This is in fact
the case: if \(a_{1}\), \(a_{2}\), \(b_{1}\), or \(b_{2}\) deviates
significantly from one, the neutrino spectrum cannot satisfy the
constraints in~(\ref{eq:NeutLimits}).  It is therefore appropriate,
since the value of \(f\) is unimportant as far as this set of
constrains are concerned (any rescaling of \(f\) can be compensated
for by a similar rescaling of \(\langle\chi\rangle\)), to analyze
constrained hierarchical models as functions of $a_{3}$ and $b_{3}$
alone.

\eit

\indent

In fig.~\ref{fig:NeutrinoMassPlots}, we show the region of viability
in \(a_{3}-b_{3}\) space for two different values of \(\delta\): in
the left-hand panel, we set \(\delta=m_{e}/m_{\mu}\) (and
\(M_{\Phi_{2}}/M_{\Phi_{3}}=m_{\mu}/m_{\tau}\)) as required in the
minimal version of CHDL discussed above; in the right-hand panel, we
set \(\delta=10^{-1}\). We consider a given combination of \(a_{3}\)
and \(b_{3}\) to be phenomenologically viable if there is any
combination of the remaining \(a_{i}\) and \(b_{i}\) for which the
combination simultaneously obeys all the neutrino oscillation
constraints in~(\ref{eq:NeutLimits}).  This plot demonstrates two
important features of the Yukawa matrices in CHDL: first, it is
indeed possible to satisfy the neutrino oscillation constraints for
\(\delta=m_{e}/m_{\mu}\); second, while \(b_{3}\), like most of the
other \(a_{i}\) and \(b_{i}\), is constrained to lie fairly close to
1, \(a_{3}\) can be quite large when \(\delta\) is small.

In the same figure, we also show contours for the value of
$\sin\theta_{13}$, the value of which will be measured or
constrained in future neutrino experiments. It is seen that the
value of $\sin\theta_{13}$ increases with increased $a_3$ until
reaching its maximum experimental bound.

\begin{figure}[ht!]
  \begin{center}
  \includegraphics{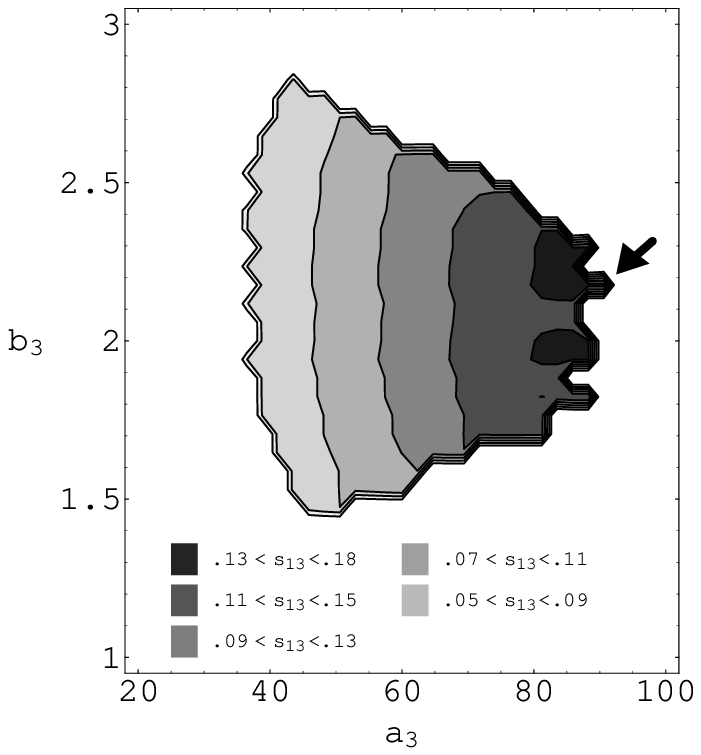}
  \includegraphics{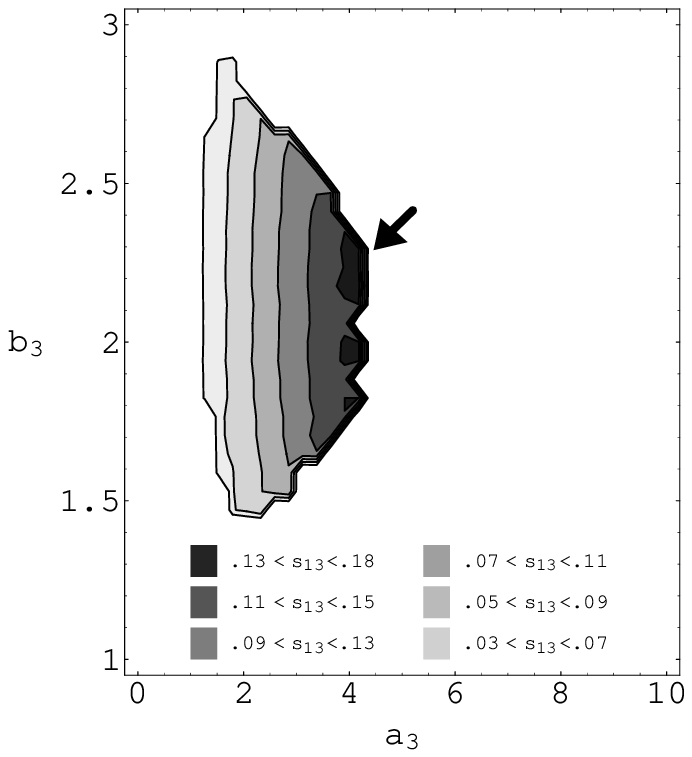}
  \end{center}
  \caption{Here the regions of \(a_{3}-b_{3}\) space (see
    equation~(\ref{eq:YukawaParametrization}) for a description of the
    Yukawa-matrix parametrization used) for which all
    constraints on neutrino masses and mixings~(\ref{eq:NeutLimits}) are simultaneously satisfied for some
    combination of the remaining \(a_{i}\) and \(b_{i}\) are shown for two different values
    of \(\delta\).  Additionally, contours depicting the ranges for \(s_{13}=\sin\theta_{13}\) (which depends primarily on
    \(a_{3}\), but varies slightly with the remaining
    \(a_{i}\) and \(b_{i}\)) are shown.  In the left-hand panel,
    \(\delta\equiv M_{\Phi_{1}}/M_{\Phi_{2}}=m_{e}/m_{\mu}\);
    in the right-hand panel, \(\delta=10^{-1}\).  The plots reveal that while
    \(b_{3}\) is constrained, along with most of the other \(a_{i}\) and \(b_{i}\), to lie
    reasonably near 1, \(a_{3}^{\mathit{max}}\) is on the order of \(10^{2}\),
    which, when \(a_{3}\approx a_{3}^{\mathit{max}}\),
    results in an increased \(\epsilon\) and---at least in the
    drift-and-decay limit---the baryon-to-photon ratio \(\eta\), making it
    easier to achieve a realistic value for \(\eta\).  For smaller
    \(\delta\), \(a_{3}^{\mathit{max}}(\delta)\) is much lower, as will
    be \(\eta\).  In each panel, the configuration that yields the
    greatest decay asymmetry \(\epsilon\) is marked with an arow.
    \label{fig:NeutrinoMassPlots}}
\end{figure}

\indent Let us now take a moment to address how these results affect
leptogenesis. Since we are assuming that \(M_{\Phi_{3}}\gg
M_{\Phi_{1}},M_{\Phi_{2}}\), the formula~(\ref{eq:epsilond}) for the
decay asymmetry \(\epsilon\) tells us that
\(\mathrm{Im}(\lambda^{\ast}_{1\alpha}\lambda_{2\alpha}h^{\ast}_{1\beta}h_{2\beta}M_{\Phi_{1}}
M^{\ast}_{\Phi_{2}})\) will vanish (when it involves diagonal
elements of \(\lambda\) or \(h\)) unless \(\beta=\alpha=3\).  This
means that
\begin{equation}
  \epsilon\propto\mathrm{Im}
  \left(a^{\ast}_{2}a_{3}b^{\ast}_{2}b_{3}\right){\delta\over 1-\delta}
\end{equation}
in CHDL; the two panels in fig.~\ref{fig:NeutrinoMassPlots} then
show that for a given \(\delta\), the largest amount of left-handed
lepton number \(L_{L}\) (and therefore the largest baryon asymmetry)
will be obtained when $a_{3}=a_{3}^{\mathit{max}}(\delta)$, where
\(a_{3}^{\mathit{max}}(\delta)\) is the maximum possible value of
\(a_{3}\) for a given $\delta$ consistent with neutrino masses and
mixings. It is interesting to note that since the maximum
experimental value of $\sin\theta_{13}$ sets the value of
$a_{3}^{\mathit{max}}$, then the maximum baryon asymmetry will be
obtained when $\sin\theta_{13}$ acquires its maximal experimental
value, that we take to be $\sin_{max}^2\theta_{13}=0.031$. Because
\(\delta=m_{e}/m_{\mu}\simeq 4.83\times10^{-3}\) is fixed to be
quite small and since \(\epsilon\) is approximately proportional to
\(\delta\) for small \(\delta\), one might worry that such a small
\(\delta\) might doom baryogenesis.  However, as indicated in the
left panel of fig.~\ref{fig:NeutrinoMassPlots},
\(a_{3}^{\mathit{max}}(m_{e}/m_{\mu})\approx 95\), so that a
hierarchy between couplings to different sets of \(\Phi\) and
\(\overline{\Phi}\) is permitted, and the result is that
\(\epsilon\) is only suppressed by an \(\mathcal{O}(1)\) numerical
factor. The right panel of fig.~\ref{fig:NeutrinoMassPlots} shows
that \(a_{3}^{\mathit{max}}(\delta)\) drops sharply as \(\delta\)
increases and \(\mathcal{O}(\delta)\) corrections become important
in \(m_{\nu}^{2}\), since \(a_{3}^{\mathit{max}}(1/10)\approx4.5\).

We are also assuming that we have maximal CP violation in the decays
of the fields $\Phi$ and $\overline{\Phi}$ (i.e. the overall phase
in the product $ a^{\ast}_{2}a_{3}b^{\ast}_{2}b_{3}$ must be
$\pi/2$). In this case one can obtain a value for the effective CP
violation in the lepton sector. This can be defined in a phase
invariant way in terms of the quantity $J={\rm
Im}(U_{12}U^*_{22}U_{23}U^*_{13})$
\cite{Jarlskog:1985ht,Dunietz:1985uy}, where $U_{ij} \equiv
(U_{\mathit{MNS}})_{ij}$. Taking for example the two points that
will generate the most baryon number in the left and right handed
panels of fig.~\ref{fig:NeutrinoMassPlots} (marked with an arrow) we
find $J\simeq0.034$ for the point in the left panel and
$J\simeq0.030$ for the point in the right panel (since the maximal
value for \(J\) is \(|U_{12}U^*_{22}U_{23}U^*_{13}|\), which for
these points yields a value of \(J\simeq.038\), the \(CP\)-violation
here is close to maximal). It could be interesting to do a more
detailed study of the issue of linking more generally the effective
CP violation in the lepton sector to the CP violation in the
interactions of the heavy fields $\Phi$ and $\overline{\Phi}$.
Nevertheless, we will not pursue this issue further in the present
work.

\indent To summarize this section, we have shown that it is possible
to construct a variety of Dirac neutrinogenesis scenarios (including
the simple, theoretically motivated CHDL) that are capable of
satisfying current constraints on the light neutrino spectrum.  In
CHDL, which will become our primary focus from this point forward,
the set of parameters relevant to neutrino physics and to
leptogenesis is quite small, given the antisymmetry condition on the
two coupling matrices $\lambda$ and $h$. The other important
parameter is the ratio of masses $\delta=M_{\Phi_{1}}/M_{\Phi_2}$.
It remains to be seen whether CHDL---or indeed any Dirac
neutrinogenesis---model can also simultaneously satisfy the
constraints from gravitino physics and yield a realistic baryon
number for the universe.  As we shall see in the next section,
thermal Dirac neutrinogenesis will indeed turn out to be workable in
a variety of split-supersymmetry models.


\section{Boltzmann Equations\label{sec:BoltzmannEquations}}

\indent

Let us now turn to the numerical calculation of \(\eta\) and the
solution of the full Boltzmann equations.
The heavy fields aside, in Dirac neutrinogenesis there are six
particle species charged under lepton number (\(\nu_{R}\),
\(\tilde{\nu}_{R}\), \(\ell\), \(\tilde{\ell}\), and the
right-handed charged lepton and slepton fields \(e_{R}\) and
\(\tilde{e}_{R}\)), and thus six individual stores of lepton number
to keep track of: \(L_{\ell}\), \(L_{\nu_{R}}\), \(L_{\tilde{L}}\),
\(L_{\tilde{\nu}_{R}}\), \(L_{e_{R}}\), and \(L_{\tilde{e}_{R}}\).
However, since \(\ell\), \(\tilde{\ell}\), \(e_{R}\), and
\(\tilde{e}_{R}\) participate in \(SU(2)\) and/or \(U(1)_{Y}\) gauge
interactions, which we assume to be sufficiently rapid (compared to
other processes relevant to leptogenesis) that these species will
always be in chemical equilibrium with one another, any lepton
number stored in any one of them will be rapidly distributed among
\(L_{\ell}\), \(L_{\tilde{\ell}}\), \(L_{e_{R}}\), and
\(L_{\tilde{e}_{R}}\) in proportion to the relative number of
degrees of freedom of each respective field.  Furthermore, since we
are assuming that \(\tilde{\ell}\) and \(\tilde{\nu}_{R}\)
equilibrate rapidly through a large \(\langle F_{\chi}\rangle\)
term, \(\tilde{\nu}_{R}\) will also be in equilibrium with these
fields. Thus \(\tilde{\nu}_{R}\), \(\ell\), \(\tilde{\ell}\),
\(e_{R}\), and \(\tilde{e}_{R}\) can be viewed as one lepton sector
with an aggregate lepton number \(L_{\mathit{agg}}\).  This leaves
only the right-handed lepton field \(\nu_{R}\), which does not
participate in any of the aforementioned rapid interactions and thus
retains its own distinct lepton number \(L_{\nu_{R}}\).

\indent  In the limit of rapid equilibration within the
\(L_{\mathit{agg}}\) sector, the set of Boltzmann equations
governing the evolution of lepton and baryon number simplifies
considerably. In addition to the relations that describe the
dynamics of the heavy fields, only three equations are required: one
for \(L_{\mathit{agg}}\), one for \(L_{\nu_{R}}\), and one for
\(B\). However, before we state these equations (a derivation of
which is provided in Appendix~\ref{app:appendix}), it will be useful
to digress for a moment and discuss the rates for the various
processes involved.  We then return to a discussion of the Boltzmann
equations themselves, followed by a presentation of the results from
our numerical computations.

\indent  While there are a large number of \(2\leftrightarrow2\)
processes which transfer lepton number between different particle
species, we will only need to calculate the rates for those which
transfer it between \(L_{\mathit{agg}}\) and \(L_{\nu_{R}}\).  As
for the rest, which collectively serve to equilibrate lepton number
among the fields in the \(L_{\mathit{agg}}\) sector, it only matters
that they are rapid compared to other rates in the problem.  The
diagrams for the \textit{s-channel} \(2\leftrightarrow2\) processes
which shuffle lepton number between \(L_{\mathit{agg}}\) and
\(L_{\nu_{R}}\) are pictured in fig.~\ref{fig:2to2SleptonProc}.  In
addition to these, there are contributions from the
\textit{t-channel} transforms (two per diagram) of these diagrams.
In order to evaluate diagrams containing virtual heavy fermions, we
define a Dirac spinor

\begin{equation}
    \Psi_{Di}=\left(\begin{array}{c} (\psi_{\Phi_{i}})_{\alpha} \\ (\psi_{\overline{\Phi}_{i}})^{\dagger\dot{\alpha}}
    \end{array}\right),
\end{equation}
where \((\psi_{\Phi_{i}})_{\alpha}\) and
\((\psi_{\overline{\Phi}_{i}})_{\alpha}\) are the Weyl spinor
components of the \(\Phi_{i}\) and \(\overline{\Phi}_{i}\)
superfields.  Numerical calculation of the thermally averaged
cross-sections for the diagram pictured on the left in the top row
of fig.~\ref{fig:2to2SleptonProc}, which involves two Yukawa-type
couplings of a scalar to two fermions, yields \(\langle\sigma
v\rangle_{i\alpha\beta}\simeq\sigma_{i}^{\mathit{(2Y)}}|\lambda_{i\alpha}|^{2}|\lambda_{i\beta}|^{2}\),
where

\begin{equation}
  \sigma_{i}^{\mathit{(2Y)}}\equiv10^{-2}\frac{T^{2}}{(M_{\Phi_{i}}^{2}+T^{2})^{2}}.
\end{equation}
For the diagram on the right in the first row of
fig.~\ref{fig:2to2SleptonProc}, which include one Yukawa-type
coupling and one trilinear scalar coupling proportional to
\(M_{\Phi_{i}}\), the result is very nearly temperature independent
and well approximated by \(\langle\sigma
v\rangle_{i\alpha\beta}\simeq\sigma_{i}^{\mathit{(1Y1S)}}|\lambda_{i\alpha}|^{2}|h_{i\beta}|^{2}\),
where

\begin{equation}
  \sigma_{i}^{\mathit{(1Y1S)}}\equiv0.5\times\frac{1}{M^{2}_{\Phi_{i}}}.
  \label{eq:1Y1S}
\end{equation}
The two diagrams in the second row of
fig.~\ref{fig:2to2SleptonProc}, which involve two Yukawa couplings
and a mass insertion from the heavy fermions, yield this same
contribution.  These interactions dominate among
\(2\leftrightarrow2\) processes.  The diagram in the bottom row of
fig.~\ref{fig:2to2SleptonProc}, which involves a trilinear scalar
coupling to the down-type Higgs, may be approximated by
\(\langle\sigma
v\rangle_{i\alpha\beta}\simeq\sigma_{i}^{\mathit{(Hd)}}|\lambda_{i\alpha}|^{2}|\lambda_{i\beta}|^{2}\),
where

\begin{equation}
  \sigma_{i}^{\mathit{(Hd)}}\propto\frac{\mu}{M^{3}_{\Phi_{i}}},
\end{equation}
where the constant of proportionality is \(\mathcal{O}(1)\), and is
thus suppressed relative to the rate given in~(\ref{eq:1Y1S}) by
\(\mu/M_{\Phi_{i}}\).  Here, we will assume that \(\mu\) is several
orders of magnitude smaller than all the \(M_{\Phi_{i}}\), and
therefore the effect of these processes can be neglected.

\begin{figure}[ht!]
\begin{center}
\includegraphics[width=8.5cm]{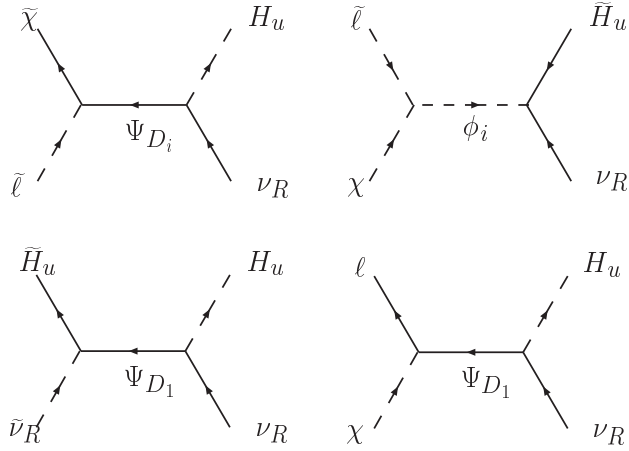}
\end{center}
  \caption{Diagrams for \(2\leftrightarrow2\) \textit{s-channel} processes which transfer lepton
    number between \(L_{\mathit{agg}}\) (the aggregate lepton number in the sector comprising the fields
    \(\ell\), \(\tilde{\ell}\), \(\tilde{\nu}_{R}\), \(e_{R}\), and
    \(\tilde{e}_{R}\), which are assumed to be in chemical equilibrium with one another due to rapid gauge and
    \(\langle F_{\chi}\rangle\)-term equilibration interactions);
     and \(L_{\nu_{R}}\) (the lepton number stored in right handed neutrinos).  The
     two \textit{t-channel}
    interactions associated with each diagram appearing above must also be included in
    calculating the full thermally averaged cross-section.
    \label{fig:2to2SleptonProc}}
\end{figure}

Taking into account contributions involving virtual fields in the
\(\Phi_{2}\) and \(\overline{\Phi}_{2}\) supermultiplets, as well as
those in \(\Phi_{1}\) and \(\overline{\Phi}_{1}\), the total
interconversion rate between \(L_{\mathit{agg}}\) and
\(L_{\nu_{R}}\) is

\begin{eqnarray}
  \Gamma_{2\leftrightarrow2} & \simeq & 3 n_{\gamma}\sum_{\alpha}\sum_{\beta}
    \left(\sigma_{1}^{\mathit{(2Y)}}|\lambda_{1\alpha}|^{2}|\lambda_{1\beta}|^{2}+\sigma_{2}^{\mathit{(2Y)}}|\lambda_{2\alpha}|^{2}|\lambda_{2\beta}|^{2}\right)\label{eq:2to2R} \non\\  & + &
    9n_{\gamma}\sum_{\alpha}\sum_{\beta}
    \left(\sigma_{1}^{\mathit{(1Y1S)}}|\lambda_{1\alpha}|^{2}|h_{1\beta}|^{2}+\sigma_{2}^{\mathit{(1Y1S)}}|\lambda_{2\alpha}|^{2}|h_{2\beta}|^{2}\right)\nonumber \\ & + &
    3n_{\gamma}\sum_{\alpha}\sum_{\beta}
    \left(\sigma_{1}^{\mathit{(Hd)}}|\lambda_{1\alpha}|^{2}|\lambda_{1\beta}|^{2}+\sigma_{2}^{\mathit{(Hd)}}|\lambda_{2\alpha}|^{2}|\lambda_{2\beta}|^{2}\right)
\end{eqnarray}
where we have assumed an equilibrium number density for all
non-leptonic light species (e.g.\ \(H_{u}\), \(\chi\)) involved.

\indent In addition to the \(2\leftrightarrow2\) processes acting,
we must also discuss the rate associated with sphaleron processes,
which represent saddle-point transitions between distinct
electroweak vacua possessing different values of \(B\) and
\(L_{\mathit{agg}}\), for these will be responsible for the
conversion of lepton number to baryon number.  As these processes
are mediated by the \(SU(2)\) electroweak anomaly, they only affect
\(L_{\ell}\) (and hence \(L_{\mathit{agg}}\)) and not
\(L_{\nu_{R}}\). We will primarily be concerned with the high
temperature sphaleron interaction rate, by which we mean the
interaction rate at temperatures \(T\gg T_{c}\), where \(T_{c}\) is
the temperature at the weak scale. This rate takes to form

\begin{equation}
  \Gamma_{\mathit{Sph}}=cT\alpha_{2}^{4},
\end{equation}
where \(\alpha_{2}\equiv g_{2}^{2}/4\pi\approx1/30\) and \(c\) is an
\(\mathcal{O}(1)\) constant, whose value
calculations~\cite{Kuzmin:1985mm,Bochkarev:1987wf,Moore:1999fs}
place very close to 1.

\indent We are now ready to write the full set of Boltzmann
equations governing the evolution of baryon number \(B\) in the
early universe.  We will need to include include equations for
\(L_{\mathit{agg}}\), \(L_{\nu_{R}}\), \(B\), as well as ones for
the lepton numbers stored in the heavy fields \(\phi_{1}\) and
\(\overline{\phi}_{1}\), which we respectively dub
\(L_{\phi_{\Phi}}\) and \(L_{\phi_{\overline{\Phi}}}\), and the
number-density-to-entropy ratio of one of these heavy fields or
their conjugates (we choose \(\phi^{c}\)), which we call
\(Y^{c}_{\phi_{\Phi}}\).  In terms of the variable $z\equiv
M_{\Phi_{1}}/T$, they are

\bea
 {dB \over dz} & = & \frac{z}{H(M_{\Phi_{1}})}\left[ -\langle\Gamma_{\mathit{sph}}\rangle
 (B+\frac{8}{15}L_{\mathit{agg}})\right] \label{eq:BoltzB}\\
 {d L_{\mathit{agg}}\over dz}&=& 
\frac{z}{H(M_{\Phi_{1}})}\left[\rule[0pt]{0pt}{16pt}-2\epsilon\langle
\Gamma_{D}\rangle(Y_{\phiphi}^c-Y_{\phiphi}^{{eq}})+\langle
\Gamma_L\rangle (L_{\phiphi} + L_{\phiphibar})\right.\non\\
&& \hspace{1.5cm} \left.+\langle\Gamma_R\rangle L_{\phiphibar}- 2
L_{\mathit{agg}}\Big(\langle \Gamma_{D}\rangle_{{}_{ID}}+\langle
\Gamma_L\rangle_{{}_{ID}}\Big)\right.\non\\
&& \hspace{1.5cm} \left. + (L_{\nu_{R}} -{1\over7}L_{\mathit{agg}})
\langle \Gamma_{2\leftrightarrow2}\rangle-
\langle\Gamma_{\mathit{sph}}\rangle
 (B+\frac{8}{15}L_{\mathit{agg}})\right] \label{eq:BoltzLnet}\\
{dL_{\nu_{R}}\over dz}&=& \frac{z}{H(M_{\Phi_{1}})}\left[2\epsilon
\langle\Gamma_{D}\rangle(Y_{\phiphi}^c-Y_{\phiphi}^{eq}) +
L_{\phiphi} \langle\Gamma_R\rangle - 2 L_{\nu_{R}} \langle
\Gamma_R\rangle_{{}_{ID}}\right.\non \\ & & \hspace{1.5cm}
\left.-(L_{\nu_{R}}
-{1\over7}L_{\mathit{agg}}) \langle \Gamma_{2\leftrightarrow2}\rangle\right] \label{eq:BoltzR}\\
{dY_{\phiphi}^c \over dz} &=& \frac{z}{H(M_{\Phi_{1}})}
\left[-\langle\Gamma_{D}\rangle(Y_{\phiphi}^c - Y_{\phiphi}^{eq})
+{1\over2} L_{\mathit{agg}}
\langle\Gamma_L\rangle_{{}_{ID}}+{1\over2} L_{\nu_{R}}\ \langle\Gamma_R\rangle_{{}_{ID}}\right] \label{eq:BoltzPhiconj}\\
{dL_{\phiphi} \over dz} &=& \frac{z}{H(M_{\Phi_{1}})}\left[-\langle
\Gamma_{D}\rangle L_{\phiphi} +2 L_{\mathit{agg}} \langle
\Gamma_L\rangle_{{}_{ID}}+2 L_{\nu_{R}}
\langle \Gamma_R\rangle_{{}_{ID}}\right]\label{eq:BoltzLPhi}\\
{dL_{\phiphibar} \over dz} &=& \frac{z}{H(M_{\Phi_{1}})}\left[
-\langle \Gamma_{D}\rangle
 L_{\phiphibar} +2 L_{\mathit{agg}}\langle
 \Gamma_{D}\rangle_{{}_{ID}}\label{eq:BoltzLPhibar}\right].
\eea Here, the inverse decay rates
\(\langle\Gamma_D\rangle_{{}_{ID}}\),
\(\langle\Gamma_L\rangle_{{}_{ID}}\), and
\(\langle\Gamma_R\rangle_{{}_{ID}}\) are defined by
\begin{eqnarray}
\langle\Gamma_D\rangle_{{}_{ID}}={1\over7}{n_{\phiphi}^{eq}\over
n_\gamma}\left({K_1(z)\over K_2(z)}\right)\Gamma_D,\\
\langle\Gamma_L\rangle_{{}_{ID}}={1\over7}{n_{\phiphi}^{eq}\over
n_\gamma}\left({K_1(z)\over K_2(z)}\right)\Gamma_L, & \mathrm{and} & \\
\langle\Gamma_R\rangle_{{}_{ID}}={n_{\phiphi}^{eq}\over
n_\gamma}\left({K_1(z)\over K_2(z)}\right)\Gamma_R,
\end{eqnarray}
where \(\Gamma_{D}\) is the total decay width of \(\phi_{1}\) given
in equation~(\ref{eq:GammaD}), \(n_{\phi}^{\mathit{eq}}\) is the
equilibrium number density of \(\phi_{1}\) (and
\(\overline{\phi}_{1}\), etc.), and the quantities \(\Gamma_{L}\)
and \(\Gamma_{R}\) represent the partial decay widths for
\(\phi\to\nu_{R}\tilde{H}\) (or
\(\overline{\phi}\to\tilde{\nu}_{R}H\)) and
\(\overline{\phi}\to\ell\tilde{\chi}\) (or
\(\phi\to\tilde{\ell}\chi\)).  The ratio of modified Bessel
functions \(K_{1}(z)\) and \(K_{2}(z)\) appearing in these rates is
a result of averaging over time-dilation factors
(see~(\ref{eq:timedilation}) in Appendix~\ref{app:appendix}).
Explicit definitions for these quantities, along with a derivation
of the Boltzmann equations themselves, are provided in
Appendix~\ref{app:appendix}.

\indent We solve this system of equations numerically as a function
of the input parameters \(\delta\) and the reparameterized coupling
strength \(f\), defined in eq.~(\ref{eq:YukawaParametrization}), and
present our results in fig.~\ref{fig:EtaContoursCCwithMRatios}.
Here, the regions of \(f-\delta\) parameter space in which the final
value of \(\eta\) generated falls within the WMAP-allowed range
given in~(\ref{eq:WMAPeta}), appear as thin `ribbons', each
corresponding to a different value of \(M_{\Phi_{1}}\). In
fig.~\ref{fig:EtaContoursCCwithMRatios}, the effects of the
processes detailed above are apparent, and certainly nontrivial.
Physically, these effects can be interpreted as follows: increasing
the strength of the neutrino-sector couplings (here parameterized by
\({f}\)) increases \(\Gamma_{D}\), which in turn increases the
initial value of \(B\); however, from equation~(\ref{eq:2to2R}),
increasing \({f}\) also increases the rates for the
\(2\leftrightarrow2\) processes which shuffle lepton number back and
forth between \(L_{\mathit{agg}}\) and \(L_{R}\).  Furthermore, it
increases the rate for inverse decays.  This allows two
possibilities for generating a realistic final value for \(B\).  In
the first case, where \({f}\) is small, the initial baryon number
produced by \(\phi\) and \(\overline{\phi}\) decays is approximately
within the range allowed by WMAP, and \(2\leftrightarrow2\) and
inverse decay processes are so slow as to be negligible; this is the
``drift-and-decay limit" of equation~(\ref{eq:DriftAndDecay}).  In
the second case, where \({f}\) is large, a surfeit of baryon number
will initially be produced, but these processes, which occur more
rapidly for larger \({f}\), subsequently reduce \(B\) to a
phenomenologically acceptable level; this we refer to as the
"strong-washout regime".  These two possibilities are shown in the
two panels of fig.~\ref{fig:SpeciesEvolution}, in which the
dynamical evolution of \(B\), \(L_{\mathit{agg}}\) and other
relevant quantities has been plotted, for
\(\delta=4.83\times10^{-3}\) and \(M_{\Phi}=10^{12}\) GeV.
In fig.~\ref{fig:EtaContoursCCwithMRatios}, the two regimes are
represented respectively by the lower and upper portions of each
ribbon.

\begin{figure}[t!]
  \begin{center}
\hspace{.25cm}
  \includegraphics{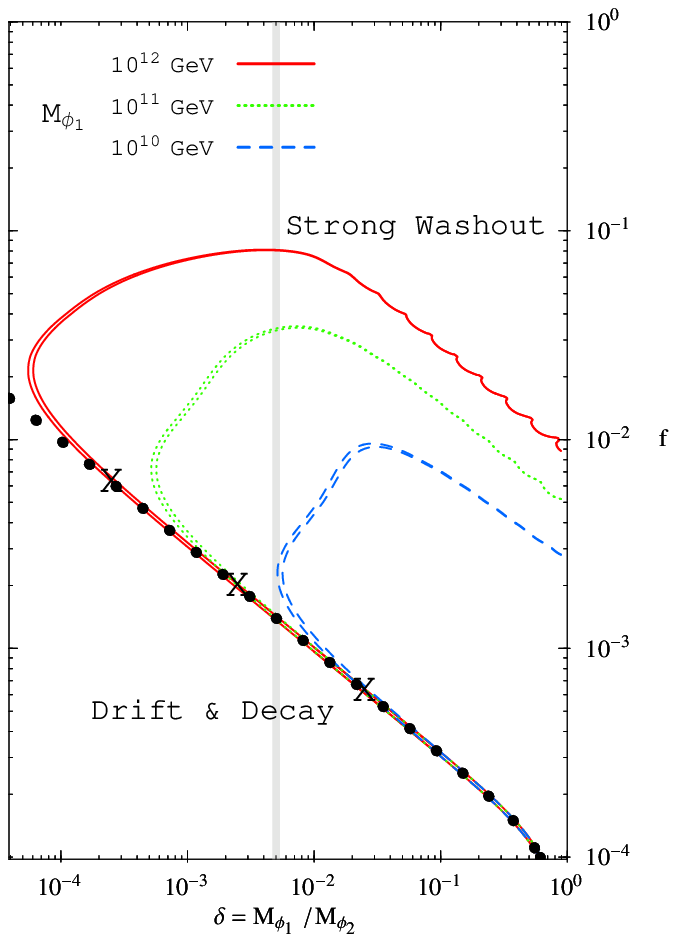}
\hspace{.5cm}
  \includegraphics{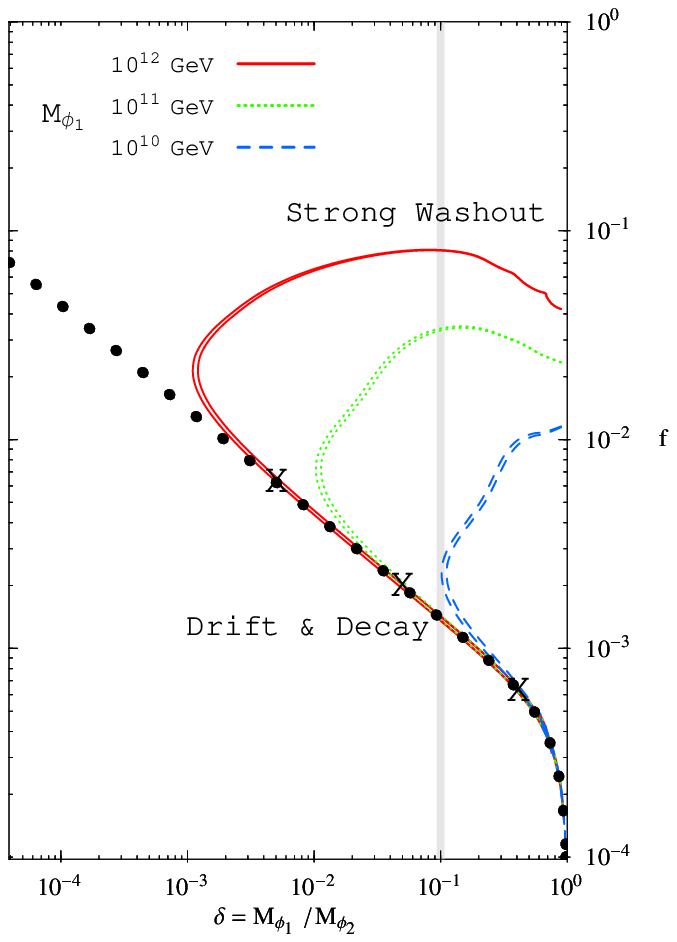}
  \end{center}
\vspace{-1cm}
  \caption{Bands in $f-\delta$ parameter space for which the final baryon
  number \(B_{f}\) 
falls within the range permitted by WMAP, for different
  choices of \(M_{\Phi_{1}}\). Here $f$ 
parameterizes the couplings of the $\Phi_i$ fields (see eq.~(\ref{eq:YukawaParametrization})).
The configuration of the left (right) panel is the one marked with a
dot in the left (right) panel of fig.~\ref{fig:NeutrinoMassPlots},
in which \(a_{3}=95\) (\(a_{3}=4.5\)). The shaded vertical lines
show the constraint on the value of $\delta$ coming from neutrino
mixings and masses (\(\delta=m_{e}/m_{\mu}\approx4.83\times10^{-3}\)
in the left and $\delta=10^{-1}$ in the right).
  Note that there are two points that yield a realistic value of \(B_{f}\) for a given \(\delta\) (the points at which the grey vertical line intersects a given ribbon).
  When \(f\) is small enough, the initial baryon
  number generated is just enough to be consistent with WMAP, while the washout effect
  of inverse decays and \(2\leftrightarrow2\) processes is negligible. In
  this situation the baryon number generated is independent
  of $M_\Phi$ and its the final value is proportional to the CP
violating parameter $\epsilon$ (see eq.~(\ref{eq:DriftAndDecay})). The dark dotted curve
  shows the band of consistent baryon number calculated in this ``drift
  and decay'' limit. Each ``X'' marks the point in which
  ${\Gamma_D/ H}=1$ for each different $M_{\Phi_1}$.
At these points, the ``strong washout'' regime starts. The now active
  washout processes reduce an initial surfeit of baryon number (due to
  a larger $f$) down to an acceptable level (see fig.~\ref{fig:SpeciesEvolution}).}
  \label{fig:EtaContoursCCwithMRatios}
\end{figure}

\indent  Having discussed the general effects of the washout
processes described above as a group, it is also important to
address their characteristics relative to one another. Inverse
decays dominate over \(2\leftrightarrow2\) processes only for a
brief period, where \(1\lesssim z \lesssim 50\), but during this
period they are extremely effective in reducing lepton number, and
in fact are the primary factor in determining the final value of
\(\eta\).  For larger \(z\), until they freeze out, the
\(2\leftrightarrow2\) interactions dominate and further reduce
\(L_{\mathit{agg}}\) and \(L_{\nu_{R}}\) (and consequently \(B\)).
It should be noted, however, that the total \(B-L_{\mathit{tot}}\)
number of the universe is manifestly conserved by the Boltzmann
equations~(\ref{eq:BoltzB}) - (\ref{eq:BoltzLPhibar}) (the sum of
the rates for the various lepton numbers involved is zero), and
since we began with \(B-L_{\mathit{tot}}=0\), we end up with
\(B-L_{\mathit{tot}}=0\) in any case.

\indent For a given \(M_{\Phi_{1}}\) and \(\delta\), then, there are
at most two points of parameter space at which all salient
constraints from neutrino physics, cosmology, and baryogenesis are
satisfied (represented in fig.~\ref{fig:EtaContoursCCwithMRatios} by
the points at which a given ribbon intersects the grey, vertical
line of constant \(\delta\)).  At each of these points, the value of
\(\langle\chi\rangle\) is obtained by requiring that the neutrino
masses given by equation~(\ref{eq:NeutMassMatrixGaugeBasis}) satisfy
the experimental bounds in equation~(\ref{eq:NeutLimits}).  For
\(M_{\Phi_{1}}=10^{12}\)~GeV and \(\delta=m_{e}/m_{\mu}\), the
strong washout case (with \(f=8.3\times10^{-2}\)) corresponds to
\(\langle\chi\rangle\sim5\)~GeV; in the drift-and-decay case (where
\(f=1.5\times10^{-3}\)), \(\langle\chi\rangle~\sim50\)~TeV.  In
general, for the strong washout regime, the value of
\(\langle\chi\rangle\) increases with decreasing \(M_{\Phi_{1}}\);
for the drift and decay case, it decreases with decreasing
\(M_{\Phi_{1}}\).  These trends continue until the two points
converge into one at \(M_{\Phi_{1}}\simeq10^{10}\)~GeV, for which
\(\langle\chi\rangle\sim1\)~TeV.

 \indent In
CHDL, the flavor structure of neutrinos rises from a hierarchy among
the $M_{\Phi_{i}}$ in which \(\delta\) is small.  As is evident from
fig.~\ref{fig:EtaContoursCCwithMRatios}, it does not seem possible
simultaneously to obtain a realistic neutrino spectrum and obtain
the correct baryon number when \(M_{\Phi_{1}}<10^{10}\) GeV.  Since
for higher \(M_{\Phi_{1}}\) we require \(T_{R}\gtrsim10^{10}\)~GeV,
the reheating temperature constraints discussed in
section~\ref{sec:AstroConstraints} are of genuine concern.  In
particular, the constraints from BBN and nonthermal LSP production
make things very difficult for CHDL when
\(m_{3/2}\lesssim10^{5}\)~GeV and BBN constraints come into play.
While it is possible to get Dirac neutrinogenesis to work when
\(10^{5}\)~GeV~\(\lesssim m_{/3/2} \lesssim 10^{8}\)~GeV, this
requires one of two things: either \(m_{\mathit{LSP}}\) is light
enough that the naive reheating temperature bound permits a
reheating temperature above \(10^{10}\)~GeV (see
fig.~\ref{fig:OmegaTot}), or else the ratio
\(m_{3/2}/m_{\mathit{LSP}}\) must be large enough that LSP
annihilations are effective (fig.~\ref{fig:OmegaAnn}).  While it is
possible to get leptogenesis to work in the former case, it is far
from optimal in the sense that nonthermal (or non-LSP) dark matter
is required.  In the latter, thermal CDM is still possible, but the
ratio \(m_{3/2}/m_{\mathit{LSP}}\) has to be be far larger than the
\(\beta/g_{\lambda}\) that results from the one-loop splitting given
in equation~(\ref{eq:AMSBMass}).  As discussed earlier, this means
there will be great difficulty getting CHDL to work in loop-split
SUSY when \(m_{3/2}\) is on the PeV scale.  There is, however, a
window where \(M_{\Phi_{1}}\sim10^{10}\)~GeV and
\(m_{LSP}\lesssim300\)~GeV in which CHDL can be made to work in such
scenarios.  For \(m_{3/2}\gtrsim10^{8}\)~GeV (where the gravitino
decays rapidly and the LSP relic abundance is entirely thermal),
reheating temperatures greater than \(10^{10}\)~GeV are permitted,
and CHDL succeeds in providing an explanation for the origin of the
observed baryon asymmetry and neutrino mixings.  Since constraints
from gluino cosmology become relevant for
\(m_{3/2}\gtrsim10^{10}\)~GeV~\cite{Arvanitaki:2005fa}, it appears
that split supersymmetry models where \(10^{8}\)~GeV~\(\lesssim
m_{3/2}\lesssim 10^{10}\)~GeV are the ``natural habitat", so to
speak, for thermal Dirac neutrinogenesis with thermal CDM.

\begin{figure}[ht!]
    \includegraphics{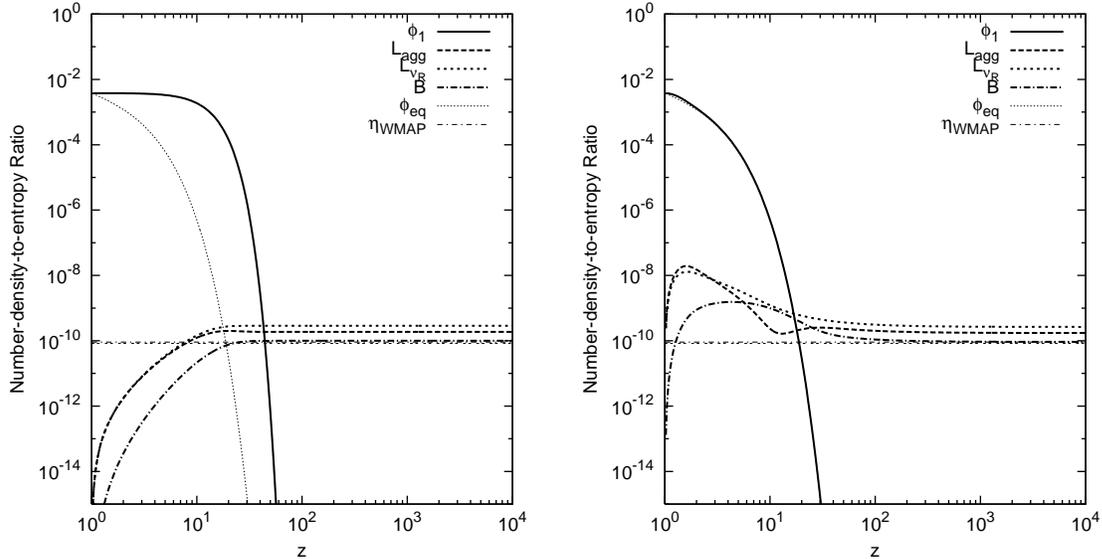}
  \caption{These two plots show the evolution of baryon number \(B\)
  for \(M_{\Phi_{1}}=10^{12}\)~GeV and
  \(\delta=m_{e}/m_{\mu}=4.83\times10^{-3}\) (the CHDL value), in the two different
  regimes that produce a realistic value for the final baryon number of the universe, \(B_{F}\).  For
  \({f}=\sqrt{\lambda_{23}h_{23}}=1.5\times10^{-3}\), as
  shown in the left panel, the effects of \(2\leftrightarrow2\)
  lepton-number-changing processes are negligible and the final
  baryon-to-entropy ratio is the same as that initially produced
  by \(\phi\) and \(\overline{\phi}\) decays.  For stronger coupling
  \({f}=3.8\times10^{-2}\), as shown in the right panel,
  baryon number is initially overproduced, but \(2\leftrightarrow2\)
  processes, which are stronger for stronger coupling, reduce \(B\)
  to an acceptable level by the time they freeze out.  The left and
  right panels correspond respectively to the lower and upper parts of the
  `ribbon' in fig.~\ref{fig:EtaContoursCCwithMRatios}.
  \label{fig:SpeciesEvolution}}
\end{figure}

\section{Conclusion\label{sec:Conclusion}}

\indent

Dirac neutrinogenesis is an interesting alternative to the standard
Majorana leptogenesis, and one which is equally capable of
explaining both the baryon asymmetry of the universe and the
smallness of neutrino masses (without the seesaw mechanism). We have
shown that thermal Dirac neutrinogenesis is not only an interesting
theory, but a genuinely viable phenomenological model: it is
simultaneously capable of producing a baryon-to-photon ratio
\(\eta\) for the universe that matches that observed by WMAP,
yielding a neutrino spectrum that agrees with the predictions of
solar and atmospheric neutrino data, and avoiding reheating
temperature constraints and other potential cosmological and
astrophysical problems. Furthermore, it is capable of doing all this
in the context of a simple setup in which the observed neutrino
masses and mixings are explained by imposing (anti)symmetry
conditions on couplings in the superpotential and a hierarchy among
the masses \(M_{\Phi_{i}}\) of the additional heavy fields in the
theory.

\indent We have also shown that Dirac neutrinogenesis is naturally
accommodated in split supersymmetry regimes involving either a heavy
gravitino (with a mass above around \(10^{8}\)~GeV) or a very large
hierarchy between the gravitino mass and the mass of the LSP.  In
models with a lighter gravitino and only a small splitting between
\(m_{3/2}\) and \(m_{LSP}\), LSP annihilations are generally
ineffective, the reheat temperature \(T_{R}\) associated with
inflation cannot be raised above \(\sim10^{10}\)~GeV, and
constraints on \(M_{\Phi_{1}}\) become quite severe.  As a result,
leptogenesis becomes extremely difficult (for ease of reference, we
classify the different gravitino-mass regimes with regard to
leptogenesis considerations in table~\ref{tab:M32Regimes}). It is
interesting to note that Dirac neutrinogenesis, at least in its
constrained hierarchical form (CHDL), can be realized in simple,
one-loop AMSB in a small region of parameter space in which
\(M_{\Phi_{1}}\sim10^{10}\)~GeV and \(m_{LSP}\lesssim300\)~GeV
(which sets \(m_{3/2}\sim10^{5}\)~GeV).  In more general split SUSY
models, cosmological constraints are less severe.  In models where
\(m_{3/2}\gtrsim10^{8}\)~GeV, the constraints altogether disappear
and Dirac neutrinogenesis can be made to work without sacrificing
the possibility for thermal LSP cold dark matter. Thus Dirac
neutrinogenesis, at least in a regime with heavy gravitinos,
provides an interesting and viable alternative to Majorana
leptogenesis.

\begin{table}[ht!]
\begin{center}
\small{\begin{tabular}{|c|c|c|l|} \hline
\parbox{2.5cm}{\begin{center}Gravitino Mass Range (GeV)\end{center}} &
\parbox{2.0cm}{\begin{center}Maximum \(M_{\Phi_{1}}\) (GeV)\end{center}} & \parbox{2.0cm}{\begin{center}Workability of CHDL?\end{center}} & Comments \\
\hline \tiny \(m_{3/2}\lesssim 10^{5}\)&  \tiny \(10^{6} - 10^{8}\)
& \tiny Very Low &
\parbox{4.5cm}{\begin{flushleft}\tiny\(\tilde{G}\) decay during or after BBN.  Insufficient \(\eta\)
generated.\end{flushleft}}\\
\hline \tiny \(10^{5}\lesssim m_{3/2}\lesssim 10^{8}\) & \tiny
\parbox{1.5cm}{\begin{center}\tiny \(10^{9} - 10^{10}\) or higher \end{center}} &
\parbox{2.0cm}{\begin{center}\tiny Depends on \(m_{\mathit{LSP}}\) and the ratio \(m_{\mathit{LSP}}/m_{3/2}\)\end{center}} & \parbox{4.5cm}{\begin{flushleft}\tiny LSP
annihilations ineffective unless \(m_{\mathit{LSP}}/m_{3/2}\) is
small. \(T_{R}\) constrained by nonthermal LSP abundance from
\(\tilde{G}\) decay.
Loop-split SUSY works only for \(m_{3/2}\sim10^{5}\)~GeV.  More general split SUSY theories can be successful. \end{flushleft}}\\
\hline \tiny \(10^{8}\lesssim m_{3/2}\lesssim 10^{10}\) & \tiny None
& \tiny Excellent &
\parbox{4.5cm}{\begin{flushleft}\tiny \(\Omega_{LSP}\) is thermal, since \(\tilde{G}\) decays
before LSP freeze-out. \(M_{\Phi_{1}}>10^{11}\)~GeV
allowed. \(\nu\) spectrum requirements compatible with CHDL.\end{flushleft}}\\
\hline \tiny \(10^{10}\lesssim m_{3/2}\lesssim 10^{12}\) & \tiny
None &
\parbox{2.0cm}{\begin{center}\tiny Questionable (depends on gluino properties)\end{center}} &
\parbox{4.5cm}{\begin{flushleft}\tiny \(\nu\) sector and baryogenesis
okay, but model may have a cosmological gluino problem.\end{flushleft}}\\
\hline \tiny \(  10^{12}\lesssim m_{3/2}\) & \tiny None & \tiny Very
Low &
\parbox{4.5cm}{\begin{flushleft}\tiny Potential gluino problem becomes a serious concern.\end{flushleft}} \\ \hline
\end{tabular}}
\end{center}
\caption{The various gravitino mass regimes for split supersymmetry models and the viability
  of thermal Constrained Hierarchical Dirac Leptogenesis (CHDL) in each case.\label{tab:M32Regimes}}
\end{table}

\indent Another asset of the theory is that there are experimental
checks on its viability. The major prediction of Dirac
neutrinogenesis is that neutrinoless double-beta decay will not be
observed to any degree, for this process relies on the existence of
a Majorana mass term for right-handed neutrinos.  The discovery of
such a process experimentally would rule the theory out.  In
addition, forthcoming results from MiniBooNE should either confirm
or deny the LSND result, which will reveal whether or not the
neutrino spectrum produced by Dirac neutrinogenesis is in fact the
one present in nature. Finally, thermal Dirac neutrinogenesis yields
some definite predictions about the neutrino mixing parameter
\(\sin\theta_{13}\) (see fig.~\ref{fig:NeutrinoMassPlots}), the
value of which will be measured in future experiments.

\section{Acknowledgments}

\indent

We would like to thank James Wells for his useful comments and
discussions and for carefully reading the manuscript. We also thank
Jason Kumar, David Morrissey, Aaron Pierce, Kazuhiro Tobe, Liliana
Velasco-Sevilla and Ting Wang for useful comments or related
discussions.  B.T. and M.T. are supported by D.O.E. and the Michigan
Center for Theoretical Physics (MCTP).


\appendix
\section{Derivation of the Boltzmann Equations\label{app:appendix}}

\indent

In this appendix, we derive the Boltzmann Equations that appear in
equations~(\ref{eq:BoltzLnet}) -- (\ref{eq:BoltzB}), following the
methods of~\cite{Kolb:1979qa}.
The Boltzmann equations for any particle species \(a\) in the early
universe can be written in terms of \(Y_{a}\equiv n_{a}/s\), where
\(n_a\) is the number density of \(a\) and \(s\) is the entropy
density, as~\cite{Kolb:1990vq}

\begin{eqnarray}
\frac{dY_{A}}{dt} &=&{1\over s} \int {d^{3}p_{a}\over (2\pi)^{3}}
{d^{3}p_{i}\over (2\pi)^{3}} {d^{3}p_{j}\over
(2\pi)^{3}}...{d^{3}p_{k}\over (2\pi)^{3}}
(2\pi)^{4}\delta(\sum_{n=a,i,j,...k}p_{n})\non \\
& & \left[\sum_{\mathit{int.}}|\mathcal{M}(a...i \to
j...k)|^{2}(f_{a}...f_{i})-\sum_{\mathit{int.}}|\mathcal{M}(j...k
\to a...i)|^{2}(f_{j}...f_{k})\right], \label{eq:DefBoltz}
\end{eqnarray}
where \(f_{i}\) is the phase-space distribution function of particle
\(i\), \(|\mathcal{M}(a...i\rightarrow j...k)|^{2}\) are the squared
matrix elements for particle-number-changing interactions involving
\(a\), and the sums are over all interaction processes which create
or destroy \(a\). Since for the scalar fields \(\phi\) and
\(\phi^{c}\), the leading processes include only decays and inverse
decays, the Boltzmann equations for the abundances \(Y_{\phi}\) and
\(Y_{\overline{\phi}}\) are given by \bea {dY_{\phiphi} \over dt}
&=&-{1\over s} \Lambda^{{}^{\phiphi}}_{12}\left[ f_{\phiphi}\ |{\cal
M}(\phiphi\to\tilde{\ell}\phichi)|^2
+f_{\phiphi}\ |{\cal M}(\phiphi\to\nu_R^c \widetilde{H}_u^c)|^2 \right.\non\\
&&-\left. f_{\phichi}f_{\widetilde{\ell}}\ |{\cal
M}(\widetilde{\ell}\phichi\to\phiphi)|^2 -f_{\nu_R^c}f_{
\widetilde{H}^c_u}\ |{\cal M}(\nu_R^c
\widetilde{H}^c_u\to\phiphi)|^2 \right] \label{eq:YPhiBasic} \eea
and \bea {dY_{\phiphi^c} \over dt} &=&-{1\over s}
\Lambda^{{}^{\phiphi^c}}_{12}\left[ f_{\phiphi^c}\ |{\cal
M}(\phiphi^c\to\widetilde{\ell}^c\phichi^c)|^2
+f_{\phiphi^c}\ |{\cal M}(\phiphi^c\to\nu_R \widetilde{H}_u)|^2 \right.\non\\
&&\left. -f_{\phichi^c}f_{\widetilde{\ell}^c}\ |{\cal
M}(\widetilde{\ell}^c\phichi^c\to\phiphi^c)|^2 -f_{\nu_R}f_{
\widetilde{H}_u}\ |{\cal M}(\nu_R \widetilde{H}_u\to\phiphi^c)|^2
\right],\label{eq:YPhibarBasic} \eea where
\(\Lambda^{a...i}_{j...k}\) has been used as a shorthand to denote
the appropriate phase-space integral and we use the notation $l$ and
$H_u$ to denote the usual Lepton and Higgs-up doublets of $SU(2)$.

\indent  Let us also define the particle asymmetry \(A\equiv
(Y_{a}-Y_{a}^{c})\), where \(Y_{a}\) and $Y_a^c$ are the entropy
normalized abundances of particle \(a\) and its conjugate $a^c$. For
a relativistic species $a$ (i.e. \(T\gg m_{a}\)) and as long as both
$A$ and $\mu_{a}$ are small, we have $A\simeq {n_\gamma\over 2s}g_a
(e^{2\mu_a/T}-1)$, which leads to $
  e^{\mu_{a}/T}\simeq 1+{s\over n_\gamma} {A\over g_{a}}$
where \(g_{a}\) is the number of degrees of freedom of particle
\(a\), $s$ is the entropy density, and $n_\gamma$ is the photon
density. If we apply the energy conservation relation to the inverse
decay processes of the form \((a+i\rightarrow b)\) appearing in
equations~(\ref{eq:YPhiBasic}) and~(\ref{eq:YPhibarBasic}), in which
particle \(i\) is in chemical equilibrium with other particles in
the thermal bath (i.e. \(\mu_{i}=0\)), the previous relation allows
us to write
\begin{equation}
    f_{a}\ f_{i}\simeq f_{b}^{\mathit{eq}}(1+{s\over n_\gamma}{A\over g_{a}}),
    \label{eq:feqsimplify}
\end{equation}
where \(f_{a}^{eq}\) is the equilibrium distribution of $a$, and the
negative sign occurs when \(a\) is a conjugate particle.

It will also be useful to relate the particle asymmetries of Leptons
to the net Lepton number abundance carried by each Lepton species.
We define the Lepton abundances as
\begin{eqnarray}
    L_\ell & \equiv & Y_{\ell} - Y_{\ell^{c}} \\
    L_{\nu_R} & \equiv & -( Y_{\nu_{R}} -  Y_{\nu_{R}^{c}}) \\
    L_{\widetilde{\ell}} & \equiv &  Y_{\widetilde{\ell}} -  Y_{\widetilde{\ell}^{c}}  \\
    L_{\widetilde{\nu}_R} & \equiv &-(  Y_{\widetilde{\nu}_{R}} -  Y_{\widetilde{\nu}_{R}^{c}}).
\end{eqnarray}
Relation~(\ref{eq:feqsimplify}), along with the \(CPT\) conservation
relation
\begin{equation}
 |{\cal M}(a\to ij)|^2=  |{\cal M}(i^{c}j^{c}\to a^c)|^2,
\end{equation}
allows us to simplify the Boltzmann equations for \(\phi\) and
\(\phi^{c}\) significantly. The results may be stated in terms of
$L_\ell$, $L_{\nu_R}$, $L_{\widetilde{\nu}_R}$ and
$L_{\widetilde{\ell}}$ as \bea {dY_{\phiphi} \over dt}
&=&-{1\over s}  \int{d^3p_{\phiphi}\over(2\pi)^3}\left[ (f_{\phiphi}
- f_{\phiphi}^{eq}) \Gamma_{D} - {s\over2n_\gamma} f_{\phiphi}^{eq}
\left( {L_{\widetilde{\ell}}\over2}\ \Gamma^c_L+
L_{\nu_R}\ \Gamma^c_R \right)\right] \label{eq:nphieqnew} \\
{dY_{\phiphi^c} \over dt} &=&-{1\over s}
\int{d^3p_{\phiphi}\over(2\pi)^3}\left[ (f_{\phiphi^c} -
f_{\phiphi}^{eq}) \Gamma_{D} + {s\over2n_\gamma} f_{\phiphi}^{eq}
\left( {L_{\widetilde{\ell}}\over2}\ \Gamma_L+ L_{\nu_R}\
\Gamma_R\right) \right], \label{eq:nphiceqnew} \eea where the
interaction rates \(\Gamma_{L}=\Gamma(\phi\to
\widetilde{\ell}+\chi)\) and \(\Gamma_{R}=\Gamma(\phi\to
\nu^c_R+\tilde{H}^c_u)\) are defined by the relations \bea
    \Gamma_{L}&=&\int \frac{d^{3}p_{i}}{(2\pi)^{3}}
    \frac{d^{3}p_{i}}{(2\pi)^{3}} |\mathcal{M}(\phi\to \ell\chi)|^{2}\\
 \Gamma_{R}&=&\int \frac{d^{3}p_{i}}{(2\pi)^{3}}
    \frac{d^{3}p_{i}}{(2\pi)^{3}} |\mathcal{M}(\phi\to \nu_R^c\tilde{H}^c_u)|^{2},
\eea with $\Gamma^c_L,R$ being the rates of the conjugate processes
and \(\Gamma_{D}=\Gamma_L+\Gamma_R\) is the total decay rate for
\(\phi\), \(\overline{\phi}\), etc.\ given in~(\ref{eq:GammaD}).
Because of supersymmetry, the total decay rate of the fermion
components of the heavy supermultiplets $\Phi$ and $\overline{\Phi}$
will also be \(\Gamma_{D}\), with the same $\Gamma_L$ and $\Gamma_R$
as their partial rates.

\indent  We also define the Lepton asymmetry  $L_{\phiphi}\equiv
Y_{\phiphi}-Y_{\phiphi^c}$ associated to the new heavy leptons.
After integrating over the incoming momentum $\vec{p}_{\phi}$ we
obtain for $L_{\phiphi}$:
\begin{equation}
{dL_{\phiphi} \over dz} = -\langle \Gamma_{D}\rangle
\left(L_{\phiphi}- {s\over2n_\gamma} Y_{\phiphi}^{eq}\epsilon
\left({L_{\widetilde{\ell}}\over2}-L_{\nu_R}\right)\right) +
Y_{\phiphi}^{eq}{s\over n_\gamma}\left(
{L_{\widetilde{\ell}}\over2}\ \langle \Gamma_L\rangle +L_{\nu_R}
\langle \Gamma_R\rangle \right)\label{eq:Lphidiffeq}
\end{equation}
Here, \(\langle \Gamma_A\rangle\) is the respective decay rate
averaged over time-dilation factors \cite{Kolb:1979qa}: \bea
\langle\Gamma_A\rangle={K_1(M_\Phi/T)\over K_2(M_\Phi/T)}\ \Gamma_A
\label{eq:timedilation} \eea where $K_1(x)$ and $K_2(x)$ are
modified Bessel functions.
The Boltzmann equations for the scalar component of the
\(\overline{\Phi}\) superfield (and its conjugate), as well as the
ones for the fermion components of $\Phi$ and $\overline{\Phi}$, are
obtained in a similar manner.


\indent  We now turn to address the evolution of the Lepton number
abundance of the particle species \(\ell\), $\nu_{R}$,
$\tilde{\ell}$, and $\tilde{\nu_{R}}$,
in which we must take into account the effect of
\(2\leftrightarrow2\) processes which transfer lepton number between
\(L_\ell\), $L_{\nu_R}$, $L_{\widetilde{\ell}}$, and
$L_{\widetilde{\nu}_R}$.
We will make one important distinction among them: the rates for
interactions which shuffle lepton number between \(\ell\),
\(\tilde{\ell}\), \(\tilde{\nu}_{R}\), and the right-handed charged
lepton and sleptons fields \(e_{R}\) and \(\tilde{e}_{R}\) will be
much larger than those for the interactions which shuffle lepton
number between \(\nu_{R}\) and any of these other fields.  This is
due to the fact that \(\nu_{R}\) only interacts via the processes
pictured in figure~\ref{fig:2to2SleptonProc}, which involve a
virtual \(\phiphi\), \(\phiphibar\), etc.\; while all of the other
fields either take part in \(SU(2)\) and/or \(U(1)_{Y}\) gauge
interactions, or---in the case of \(\tilde{\nu_{R}}\)---left-right
slepton equilibration through an assumed large \(\langle
F_{\chi}\rangle\) term.
We will represent the effects of these rapid equilibration processes
by including terms \(\Sigma_{A}\) (where \(A\) is the relevant
particle asymmetry) to represent them in the Boltzmann equations.
The slower \(2\leftrightarrow2\) processes through which right
handed neutrinos \(\nu_R\) interact with lepton doublets $\ell$ and
\(\widetilde{\ell}\) and with right handed sneutrino
\(\widetilde{\nu}_R\) will be included separately as
\(C_{\nu_R\leftrightarrow \ell \vphantom{\tilde{A}}}\),
\(C_{\nu_R\leftrightarrow\widetilde{\ell}}\) and
\(C_{\nu_R\leftrightarrow\widetilde{\nu}_R}\). To simplify further
the notation we will define the terms \(F_{A}\) to account for the
collective contribution from decays (and inverse decays) of the
fermionic components of \(\Phi\) and \(\overline{\Phi}\) (which we
will not write explicitly, being of similar form to the contribution
from the scalar components).

\indent We obtain for the Lepton number abundance \(L_\ell\): \bea
{dL_\ell\over dt}
&=&-\epsilon\ \langle\Gamma_{D}\rangle
\Big(Y_{\phiphibar^c}+Y^{eq}_{\phiphi}\Big)+ L_{\phiphibar}
\langle\Gamma_L\rangle-Y^{eq}_{\phiphi}{s\over n_\gamma} L_\ell\
\Big(\langle\Gamma_L\rangle-{1\over2}\epsilon
\langle\Gamma_{D}\rangle \Big)\non\\
&&  +F_\ell+\Sigma_\ell +{C}_{\nu_R\leftrightarrow
\ell\vphantom{\widetilde{A}}} \eea Similarly, the equations for
\(L_{\nu_R}\), \(L_{\widetilde{\ell}}\), and
\(L_{\widetilde{\nu}_R}\) are given by \bea {dL_{\nu_R}\over dt}&=&
\epsilon\ \langle \Gamma_{D}\rangle
\Big(Y^c_{\phiphi}+Y^{eq}_{\phiphi}\Big) -L_{\phiphi}
\langle\Gamma_{R}\rangle + Y^{eq}_{\phiphi} {s\over n_\gamma}
L_{\nu_R} \Big(\langle\Gamma_R\rangle-{1\over2}\epsilon
\Gamma_{D})\Big)\non\\
&&  + F_{\nu_R}
-C_{\nu_R\leftrightarrow\widetilde{\ell}}-{C}_{\nu_R\leftrightarrow\widetilde{\nu}_R}-{C}_{\nu_R\leftrightarrow
\ell\vphantom{\widetilde{A}}}
\label{eq:Rdiffeq}\\
{dL_{\widetilde{\ell}}\over dt}&=& - \epsilon\ \langle
\Gamma_{D}\rangle
 (Y_{\phiphi^c}+Y^{eq}_{\phiphi})
+ L_{\phiphi}\langle\Gamma_L\rangle - Y^{eq}_{\phiphi} {s\over
n_\gamma} L_{\widetilde{\ell}}
\Big({1\over2}\langle\Gamma_L\rangle+{1\over2}\epsilon
\langle\Gamma_{D}\rangle\Big)\non\\
&&+F_{\widetilde{\ell}}+\Sigma_{\tilde{\ell}}+\ C_{\nu_R\to\widetilde{\ell}} \\
{dL_{\widetilde{\nu}_R}\over dt}&=&  \epsilon\ \langle
\Gamma_{D}\rangle (Y_{\phiphibar^c}+Y^{eq}_{\phiphibar}) -
L_{\phiphibar}\langle\Gamma_R\rangle -Y^{eq}_{\phiphibar}{s\over
n_\gamma} L_{\widetilde{\nu}_R}
\Big(\langle\Gamma_R\rangle+{1\over2}\epsilon
\Gamma_{D}\Big)\non\\
&& +F_{\widetilde{\nu}_R}+\Sigma_{\widetilde{\nu}_R}+
C_{\nu_R\to\widetilde{\nu}_R}. \eea

Before going any further we still need to compute the terms
$C_{\nu_R\leftrightarrow\widetilde{\ell}}$,
${C}_{\nu_R\leftrightarrow\widetilde{\nu}_R}$ and
${C}_{\nu_R\leftrightarrow \ell\vphantom{\widetilde{A}}}$
corresponding to the $2\leftrightarrow 2$ processes mediated by
heavy fields.

We begin by calculating \(C_{\nu_R\leftrightarrow\tilde{L}}\), which
is given by \bea C_{\nu_R\leftrightarrow\widetilde{\ell}} &=&{2\over
n_\gamma}\Lambda^{34}_{12}e^{-(E_1+E_2)/T} \left[\Big(|{\cal
M'}(\widetilde{H}_u^c \nu_R^c\to\widetilde{\ell}\chi)|^2-|{\cal
M'}(\widetilde{\ell}\chi\to\nu_R^c \widetilde{H}^c_u)|^2\Big)
\vphantom{\int_\int}\right.\non\\
&& \left.+
{(L_{\nu_R}-{1\over2}L_{\widetilde{\ell}})\over2}\Big(|{\cal
M'}(\widetilde{H}_u^c \nu_R^c\to\widetilde{\ell}\chi)|^2 +|{\cal
M'}(\widetilde{\ell}\chi\to\nu_R^c
\widetilde{H}^c_u)|^2\Big)\vphantom{\int_\int} \right]
\label{eq:CRL} \eea ${\cal M'}$ here refers to the amplitude for the
specified $2\leftrightarrow 2$ process to which we have substracted
the contribution from the resonant intermediate state (RIS) in which
a real field $\phiphi$ is produced and then decayed into the 2
particle final state. The RIS contribution must be substracted since
we have already counted contributions from decays of real $\phiphi$
fields.

\indent The leading term in the difference between a
$2\leftrightarrow 2$ process of the form \(ab\to ij\) involving a
heavy intermediary \(k\) and its conjugate process depends on the
contribution of the on-shell (resonant) intermediate state: \beq
\!\!\!|{\cal M'}(ab\to ij)|^2-|{\cal M'}(ij\to ab)|^2 =|{\cal
M_{RIS}}(ij\to ab)|^2 -|{\cal M_{RIS}}(ab\to ij)|^2 \eeq with \beq
\hspace{-.5cm}|{\cal M_{RIS}}(ab\to ij)|^2 \simeq {\pi\over m_\phi
\Gamma_{D}} \delta(s-m_\phi^2) |{\cal M}(ab\to k)|^2\times|{\cal
M}(k\to ij)|^2
\eeq where \(k\) represents the intermediate-state particle, and $s$
is the usual kinematic variable $s=(p^{in}_1+p^{in}_2)^2$.  In our
case, making use of the equality
$\Gamma_L^c\Gamma_R-\Gamma_L\Gamma_R^c=\epsilon \Gamma_{D}$ we find
\bea |{\cal M_{RIS}}(\widetilde{\ell}\chi\to\nu_R^c
\widetilde{H}^c_u)|^2 -|{\cal M_{RIS}}(\widetilde{H}^c_u
\nu_R^c\to\widetilde{\ell}\chi)|^2 \simeq \epsilon {\pi\over m_\phi
\Gamma_{D}}\delta(s-m_\phi^2) |{\cal M}^\phi_{tot}|^4 \eea and
substituting this result in eq.~(\ref{eq:CRL}) we finally obtain
\bea C_{\nu_R\leftrightarrow\widetilde{\ell}}&=& 2 \epsilon
Y_{\phiphi}^{eq} \langle \Gamma_{D}\rangle +
(L_{\nu_R}-{1\over2}L_{\widetilde{\ell}})\ n_\gamma \langle
v\sigma_{\nu_R\to \widetilde{\ell}}+v\sigma_{\widetilde{\ell}\to
\nu_R}\rangle. \eea For the other sets of \(2\leftrightarrow2\)
processes, \(C_{\nu_R\to\widetilde{\nu}_R}\) and \(C_{\nu_R\to
\ell}\), the procedure is essentially the same. The rates
\(\langle\Gamma_{\nu_R\leftrightarrow \widetilde{\ell}}\rangle\equiv
n_\gamma \langle v\sigma_{\nu_R\to
\widetilde{\ell}}+v\sigma_{\tilde{\ell}\to \nu_R}\rangle\),
\(\langle\Gamma_{\nu_R\leftrightarrow \ell}\rangle\equiv n_\gamma
\langle v\sigma_{\nu_R\to \ell}+v\sigma_{\ell\to \nu_R}\rangle\) and
\(\langle\Gamma_{\nu_R\leftrightarrow\widetilde{\nu}_R}\rangle\equiv
n_\gamma \langle v\sigma_{\nu_R\to
\widetilde{\nu}_R}+v\sigma_{\widetilde{\nu}_R\to \nu_R}\rangle\),
associated with these interactions are calculated in
section~\ref{sec:BoltzmannEquations}. We will denote the total
contribution from these processes as
\(\langle\Gamma_{2\leftrightarrow2}\rangle\).

\indent  At this point, the full Boltzmann system comprises fourteen
individual differential equations: four for \(\phiphi\),
\(\phiphibar\), and their conjugates (or alternatively, the
abundances \(Y_{\phiphi}\), \(L_{\phiphi}\), etc.); four for the the
Lepton asymmetries $L_{\ell}$, $L_{\nu_R}$, $L_{\widetilde{\ell}}$,
and \(L_{\widetilde{\nu}_R}\); two for the Lepton asymmetries
\(L_{e_R}\equiv e^c_{R}-e_{R}\) and \(L_{\widetilde{e}_R}\equiv
\widetilde{e}^c_{R}-\widetilde{e}_{R}\) of the right-handed charged
lepton and slepton fields; and an additional four for the fermionic
superpartners in the \(\Phi\) and \(\overline{\Phi}\)
supermultiplets.

Let us make the assumption that the rates for the processes denoted
by \(\Sigma_{A}\) are sufficiently rapid that chemical equilibrium
is achieved among the particle species that interact via these
processes.  In this case, any lepton number stored in \(\ell\),
\(\tilde{\ell}\), \(\tilde{\nu}_{R}\), \(e_{R}\), or
\(\tilde{e}_{R}\) should be rapidly distributed among all of these
particles in proportion to the relative number of degrees of freedom
for each field.  The lepton number stored in each field then becomes
\bea
L_\ell,L_{\widetilde{\ell}}&\to& L_{eq},\\
L_{\widetilde{\nu}_R},L_{e_R},L_{\widetilde{e}_R}&\to&L_{eq}/2, \eea
where we have defined an equilibrium lepton number
\(L_{\mathit{eq}}\) which represents the effects of rapid
equilibration through \(\Sigma_{A}\).  Because of this rapid
equilibration, it is no longer necessary to keep track of the
individual lepton numbers \(L_{\ell}\), \(L_{\widetilde{\ell}}\),
\(L_{\widetilde{\nu}_R}\), \(L_{e_R}\), and \(L_{\widetilde{e}_R}\):
we need only one differential equation representing the total lepton
number in this equilibrated, aggregate sector
$L_{\mathit{agg}}=L_\ell+L_{\widetilde{\nu}_R}+L_{\widetilde{\ell}}+L_{e_R}+L_{\widetilde{e}_R}={7\over2}
L_{eq}$. The equation is \bea {d L_{\mathit{agg}}\over dt} &=&
-\epsilon\langle \Gamma_{D}\rangle
\left(Y_{\phiphi}^c-Y_{\phiphi}^{eq}\right)+\langle \Gamma_L\rangle
\left(L_{\phiphi} + L_{\phiphibar}\right) + \langle\Gamma_R\rangle
L_{\phiphibar}\non\\
&& - Y_{\phiphi}^{eq} {s\over 2 n_\gamma} L_{\mathit{agg}}
\Big(\langle\Gamma_D\rangle +\langle\Gamma_L\rangle \Big)+
(L_{\nu_R}-{1\over7}L_{\mathit{agg}})\
\langle\Gamma_{2\leftrightarrow2}\rangle+F_{\mathit{net}}
,\label{eq:Lnetdiffeq} \eea where \(F_{\mathit{net}}\) represents
the overall contribution to \(L_{\mathit{agg}}\) from the fermionic
components of \(\Phi\) and \(\overline{\Phi}\), and small terms
proportional to $\epsilon$ times $L_\ell$, $L_{\nu_R}$,
$L_{\widetilde{\ell}}$ or $L_{\widetilde{\nu}_R}$ have been dropped.
Here, we have used the fact that the Boltzmann equations for
\(L_{e_R}\) and \(L_{\widetilde{e}_R}\) are trivial, consisting of
only \(\Sigma_{A}\) terms, which all cancel after taking the sum of
all Lepton abundances.

\indent  The one task that remains is to deal with the fermionic
components of \(\Phi\) and \(\overline{\Phi}\).  However, it can be
shown that in the approximation that the \(\Sigma_{A}\) interactions
are rapid, the decay and inverse decay contributions to
\(L_{\mathit{agg}}\) and \(L_{\nu_{R}}\) from these fermions are the
same as those from the scalars, and thus no additional Boltzmann
equations for \(\psi_{\Phi}\), \(\psi_{\overline{\Phi}}\), etc.\ are
needed. We therefore require only five differential equations to
describe the evolution of lepton number in the universe.
These are \bea {d L_{\mathit{agg}}\over dt}&=& 
-2\epsilon\langle
\Gamma_{D}\rangle(Y_{\phiphi}^c-Y_{\phiphi}^{{eq}})+\langle
\Gamma_L\rangle (L_{\phiphi} + L_{\phiphibar}) +
\langle\Gamma_R\rangle L_{\phiphibar}\non\\ && - 2
L_{\mathit{agg}}\Big(\langle \Gamma_{D}\rangle_{{}_{ID}}+\langle
\Gamma_L\rangle_{{}_{ID}}\Big)
+ (L_{\nu_{R}} -{1\over7}L_{\mathit{agg}}) \langle \Gamma_{2\leftrightarrow2}\rangle\\
{dL_{\nu_R}\over dt}&=& 2\epsilon
\langle\Gamma_{D}\rangle(Y_{\phiphi}^c-Y_{\phiphi}^{eq}) +
L_{\phiphi} \langle\Gamma_R\rangle - 2 L_{\nu_R} \langle
\Gamma_R\rangle_{{}_{ID}}\non \\ & & -(L_{\nu_{R}}
-{1\over7}L_{\mathit{agg}}) \langle \Gamma_{2\leftrightarrow2}\rangle\\
{dY_{\phiphi}^c \over dt} &=&-\langle\Gamma_{D}\rangle(Y_{\phiphi}^c
- Y_{\phiphi}^{eq}) +{1\over2} L_{\mathit{agg}}
\langle\Gamma_L\rangle_{{}_{ID}}+{1\over2} L_{\nu_{R}}\ \langle\Gamma_R\rangle_{{}_{ID}}\\
{dL_{\phiphi} \over dt} &=& -\langle \Gamma_{D}\rangle L_{\phiphi}
+2 L_{\mathit{agg}} \langle \Gamma_L\rangle_{{}_{ID}}+2 L_{\nu_{R}}
\langle \Gamma_R\rangle_{{}_{ID}}\\
{dL_{\phiphibar} \over dt} &=& -\langle \Gamma_{D}\rangle
 L_{\phiphibar} +2 L_{\mathit{agg}}\langle
 \Gamma_{D}\rangle_{{}_{ID}},
\eea where \(\displaystyle
\langle\Gamma_D\rangle_{{}_{ID}}={1\over7}{n_{\phiphi}^{eq}\over
n_\gamma}\langle\Gamma_D\rangle,\ \
\langle\Gamma_L\rangle_{{}_{ID}}={1\over7}{n_{\phiphi}^{eq}\over
n_\gamma}\langle\Gamma_L\rangle\ \) and\  \(\displaystyle
\langle\Gamma_R\rangle_{{}_{ID}}={n_{\phiphi}^{eq}\over
n_\gamma}\langle\Gamma_D\rangle\). Once again, small terms
proportional to \(\epsilon L_{\mathit{agg}}\) or \(\epsilon
L_{\nu_{R}}\) have been neglected.  Of course we have not yet
included the coupling of baryon number \(B\) to
\(L_{\mathit{agg}}\), which is accomplished by introducing a
differential equation for \(B\) (making a total of six coupled
equations) and the addition of a sphaleron interaction term which
mixes \(B\) and \(L_{\mathit{agg}}\). With this addition one obtains
equations~(\ref{eq:BoltzB}) -- (\ref{eq:BoltzLPhibar}).

Before closing this appendix we can easily solve these equations in
the ``drift and decay'' limit, in which we assume that all the rates
$\langle\Gamma_A\rangle$ are much smaller than the rate of expansion
of the universe \(H\). In that limit the equations simplify greatly:
\bea {d L_{\mathit{agg}}\over dt}&=& 
-2\epsilon\langle
\Gamma_{D}\rangle(Y_{\phiphi}^c-Y_{\phiphi}^{{eq}})\\
{dL_{\nu_R}\over dt}&=& 2\epsilon
\langle\Gamma_{D}\rangle(Y_{\phiphi}^c-Y_{\phiphi}^{eq}) \\
{dY_{\phiphi}^c \over dt} &=&-\langle\Gamma_{D}\rangle(Y_{\phiphi}^c
- Y_{\phiphi}^{eq}), \eea from which we can write \bea
{dL_{\nu_R}\over dt}=-{dL_{\mathit{agg}}\over dt}=-2\epsilon\
{dY_{\phiphi}^c \over dt}. \eea Since the inital values of
$L_{\nu_R}$ and $L_{\mathit{agg}}$ vanish, the final abundances
after a long time will be \bea
L_{\nu_R}^{final}=-L^{final}_{\mathit{agg}}
=2\epsilon Y_{\phiphi}^{eq}(t=0) \eea where we have assumed that
$Y_{\phiphi}^{final}=0 $ and $Y_{\phiphi}^{initial}=
Y_{\phiphi}^{eq}(t\!=\!0)$ with $t=0$ defined as the time in which
$T=M_\Phi$.



\begin{thebibliography}{99}

\bibitem{Arkani-Hamed:2004fb}
  N.~Arkani-Hamed and S.~Dimopoulos,
  ``Supersymmetric Unification without Low Energy Supersymmetry and  Signatures
  for Fine-tuning at the LHC,''
  JHEP {\bf 0506}, 073 (2005)
  [hep-th/0405159];
  N.~Arkani-Hamed, S.~Dimopoulos, G.~F.~Giudice and A.~Romanino,
  ``Aspects of Split Supersymmetry,''
  Nucl.\ Phys.\ B {\bf 709}, 3 (2005)
  [hep-ph/0409232].

\bibitem{Giudice:2004tc}
G.~F.~Giudice and A.~Romanino, ``Split Supersymmetry,'' Nucl.\
Phys.\ B {\bf 699}, 65 (2004) [Erratum-ibid.\ B {\bf 706}, 65
(2005)] [hep-ph/0406088].

\bibitem{Wells:2003tf}
J.~D.~Wells, ``Implications of Supersymmetry Breaking with a Little
Hierarchy Between Gauginos and Scalars,'' [hep-ph/0306127];
J.~D.~Wells, ``PeV-scale Supersymmetry,'' [hep-ph/0411041].

\bibitem{Randall:1998uk}
L.~Randall and R.~Sundrum, ``Out of This World Supersymmetry
Breaking,'' Nucl.\ Phys.\ B {\bf 557}, 79 (1999) [hep-th/9810155];
G.~F.~Giudice, M.~A.~Luty, H.~Murayama and R.~Rattazzi, ``Gaugino
Mass without Singlets,'' JHEP {\bf 9812}, 027 (1998)
[hep-ph/9810442].

\bibitem{Masiero:2004ft}
A.~Masiero, S.~Profumo and P.~Ullio, ``Neutralino Dark Matter
Detection in Split Supersymmetry Scenarios,'' [hep-ph/0412058].

\bibitem{Arvanitaki:2004df}
A.~Arvanitaki and P.~W.~Graham, ``Indirect Signals from Dark Matter
in Split Supersymmetry,'' [hep-ph/0411376].

\bibitem{Thomas:2005te}
  B.~Thomas,
  ``Requirements to Detect the Monoenergetic Photon Signature of Thermal Cold
  Dark Matter in PeV-scale Split Supersymmetry,''
  [hep-ph/0503248].

\bibitem{Toharia:2005gm}
  M.~Toharia and J.~D.~Wells,
  ``Gluino Decays with Heavier Scalar Superpartners,''
  [hep-ph/0503175].

\bibitem{Gambino:2005eh}
  P.~Gambino, G.~F.~Giudice and P.~Slavich,
  ``Gluino Decays in Split Supersymmetry,''
  Nucl.\ Phys.\ B {\bf 726}, 35 (2005)
  [hep-ph/0506214].

\bibitem{Dick:1999je}
  K.~Dick, M.~Lindner, M.~Ratz and D.~Wright,
  ``Leptogenesis with Dirac Neutrinos,''
  Phys.\ Rev.\ Lett.\  {\bf 84}, 4039 (2000)
  [hep-ph/9907562].

\bibitem{Murayama:2002je}
  H.~Murayama and A.~Pierce,
  ``Realistic Dirac Leptogenesis,''
  Phys.\ Rev.\ Lett.\  {\bf 89}, 271601 (2002)
  [hep-ph/0206177].

\bibitem{Green:1984sg}
  M.~B.~Green and J.~H.~Schwarz,
  ``Anomaly Cancellation in Supersymmetric D=10 Gauge Theory and Superstring
  Theory,''
  Phys.\ Lett.\ B {\bf 149}, 117 (1984).

\bibitem{Borzumati:2000mc}
  F.~Borzumati and Y.~Nomura,
  ``Low-scale See-saw Mechanisms for Light Neutrinos,''
  Phys.\ Rev.\ D {\bf 64}, 053005 (2001)
  [hep-ph/0007018].

\bibitem{Bennett:2003bz}
  C.~L.~Bennett {\it et al.},
  ``First Year Wilkinson Microwave Anisotropy Probe (WMAP) Observations:
  Preliminary Maps and Basic Results,''
  Astrophys.\ J.\ Suppl.\  {\bf 148}, 1 (2003)
  [astro-ph/0302207].

\bibitem{Sakharov:1967dj}
  A.~D.~Sakharov,
  ``Violation Of CP Invariance, C Asymmetry, and Baryon Asymmetry of the
  Universe,''
  Pisma Zh.\ Eksp.\ Teor.\ Fiz.\  {\bf 5}, 32 (1967)
  [JETP Lett.\  {\bf 5}, 24 (1967\ SOPUA,34,392-393.1991\ UFNAA,161,61-64.1991)].

\bibitem{Cohen:1987vi}
  A.~G.~Cohen and D.~B.~Kaplan,
  ``Thermodynamic Generation of the Baryon Asymmetry,''
  Phys.\ Lett.\ B {\bf 199}, 251 (1987).

\bibitem{'tHooft:1976fv}
  G.~'t Hooft,
  ``Computation of the Quantum Effects Due to a Four-Dimensional
  Pseudoparticle,''
  Phys.\ Rev.\ D {\bf 14}, 3432 (1976)
  [Erratum-ibid.\ D {\bf 18}, 2199 (1978)].

\bibitem{Affleck:1984fy}
  I.~Affleck and M.~Dine,
  ``A New Mechanism for Baryogenesis,''
  Nucl.\ Phys.\ B {\bf 249}, 361 (1985).

\bibitem{Fukugita:1986hr}
  M.~Fukugita and T.~Yanagida,
  ``Baryogenesis without Grand Unification,''
  Phys.\ Lett.\ B {\bf 174}, 45 (1986).

\bibitem{Luty:1992un}
  M.~A.~Luty,
  ``Baryogenesis via Leptogenesis,''
  Phys.\ Rev.\ D {\bf 45}, 455 (1992).

\bibitem{Buchmuller:2002xm}
  W.~Buchmuller,
  ``Neutrinos, Grand Unification and Leptogenesis,''
  [hep-ph/0204288]; W.~Buchmuller, P.~Di Bari and M.~Plumacher,
  ``Leptogenesis for Pedestrians,''
  Annals Phys.\  {\bf 315}, 305 (2005)
  [hep-ph/0401240];
  W.~Buchmuller, P.~Di Bari and M.~Plumacher,
  ``Some Aspects of Thermal Leptogenesis,''
  New J.\ Phys.\  {\bf 6}, 105 (2004)
  [hep-ph/0406014].

\bibitem{Yanagida:1979as}
  T.~Yanagida,
  ``Horizontal Gauge Symmetry and Masses of Neutrinos,''
{\it In Proceedings of the Workshop on the Baryon Number of the
Universe and Unified Theories, Tsukuba, Japan, 13-14 Feb 1979};
  M.~Gell-Mann, P.~Ramond and R.~Slansky,
  in {\it Supergravity} (North Holland, Amsterdam, 1979)
  eds.\ P.~Van Nieuenhuizen, D.~Freedman, p.\ 315.

\bibitem{Flanz:1996fb}
  M.~Flanz, E.~A.~Paschos, U.~Sarkar and J.~Weiss,
  ``Baryogenesis through Mixing of Heavy Majorana Neutrinos,''
  Phys.\ Lett.\ B {\bf 389}, 693 (1996)
  [hep-ph/9607310].

\bibitem{Kolb:1990vq}
  E.~W.~Kolb and M.~S.~Turner,
  ``The Early Universe,''

\bibitem{Olive:1999ij}
  K.~A.~Olive, G.~Steigman and T.~P.~Walker,
  ``Primordial Nucleosynthesis: Theory and Observations,''
  Phys.\ Rept.\  {\bf 333}, 389 (2000)
  [astro-ph/9905320].

\bibitem{Moroi:1995fs}
  T.~Moroi,
  ``Effects of the Gravitino on the Inflationary Universe,''
  arXiv:hep-ph/9503210.

\bibitem{Kawasaki:2004qu}
  M.~Kawasaki, K.~Kohri and T.~Moroi,
  ``Big-bang Nucleosynthesis and Hadronic Decay of Long-lived Massive
  Particles,''
  Phys.\ Rev.\ D {\bf 71}, 083502 (2005)
  [astro-ph/0408426].

\bibitem{Spergel:2003cb}
D.~N.~Spergel {\it et al.}  [WMAP Collaboration], ``First Year
Wilkinson Microwave Anisotropy Probe (WMAP) Observations:
Determination of Cosmological Parameters,'' Astrophys.\ J.\ Suppl.\
{\bf 148}, 175 (2003) [astro-ph/0302209].

\bibitem{Kawasaki:1994af}
  M.~Kawasaki and T.~Moroi,
  ``Gravitino Production in the Inflationary Universe and the Effects on Big
  Bang Nucleosynthesis,''
  Prog.\ Theor.\ Phys.\  {\bf 93}, 879 (1995)
  [hep-ph/9403364].

\bibitem{Ibe:2004tg}
  M.~Ibe, R.~Kitano, H.~Murayama and T.~Yanagida,
  ``Viable Supersymmetry and Leptogenesis with Anomaly Mediation,''
  Phys.\ Rev.\ D {\bf 70}, 075012 (2004)
  [hep-ph/0403198].

\bibitem{Moroi:1999zb}
  T.~Moroi and L.~Randall,
  ``Wino Cold Dark Matter from Anomaly-mediated SUSY Breaking,''
  Nucl.\ Phys.\ B {\bf 570}, 455 (2000)
  [hep-ph/9906527].

\bibitem{Allahverdi:2005rh}
  R.~Allahverdi, S.~Hannestad, A.~Jokinen, A.~Mazumdar and S.~Pascoli,
  ``Supermassive Gravitinos, Dark Matter, Leptogenesis and Flat Direction
  Baryogenesis,''
  [hep-ph/0504102].

\bibitem{Arvanitaki:2005fa}
  A.~Arvanitaki, C.~Davis, P.~W.~Graham, A.~Pierce and J.~G.~Wacker,
  ``Limits on Split Supersymmetry from Gluino Cosmology,''
  [hep-ph/0504210].

\bibitem{Hagedorn:2005kz}
  C.~Hagedorn and W.~Rodejohann,
  ``Minimal Mass Matrices for Dirac Neutrinos,''
  [hep-ph/0503143].

\bibitem{Davidson:1980sx}
  A.~Davidson and K.~C.~Wali,
  ``Symmetric Versus Antisymmetric Mass Matrices in Grand Unified Theories,''
  Phys.\ Lett.\ B {\bf 94}, 359 (1980).

\bibitem{King:2004tx}
  S.~F.~King, I.~N.~R.~Peddie, G.~G.~Ross, L.~Velasco-Sevilla and O.~Vives,
  ``Kaehler Corrections and Softly Broken Family Symmetries,''
  JHEP {\bf 0507}, 049 (2005)
  [hep-ph/0407012].

\bibitem{Froggatt:1978nt}
  C.~D.~Froggatt and H.~B.~Nielsen,
  ``Hierarchy of Quark Masses, Cabibbo Angles and CP Violation,''
  Nucl.\ Phys.\ B {\bf 147}, 277 (1979).

\bibitem{Bahcall:2001zu}
  J.~N.~Bahcall, M.~C.~Gonzalez-Garcia and C.~Pena-Garay,
  ``Global Analysis of Solar Neutrino Oscillations Including SNO CC
  Measurement,''
  JHEP {\bf 0108}, 014 (2001)
  [hep-ph/0106258].

\bibitem{Fogli:2001vr}
  G.~L.~Fogli, E.~Lisi, D.~Montanino and A.~Palazzo,
  ``Model-dependent and Independent Implications of the First Sudbury  Neutrino
  Observatory Results,''
  Phys.\ Rev.\ D {\bf 64}, 093007 (2001)
  [hep-ph/0106247].


\bibitem{Jarlskog:1985ht}
  C.~Jarlskog,
  ``Commutator of The Quark Mass Matrices in the Standard Electroweak Model and
  a Measure of Maximal CP Violation,''
  Phys.\ Rev.\ Lett.\  {\bf 55}, 1039 (1985).

\bibitem{Dunietz:1985uy}
  I.~Dunietz, O.~W.~Greenberg and D.~d.~Wu,
  ``A Priori Definition of Maximal CP Violation,''
  Phys.\ Rev.\ Lett.\  {\bf 55}, 2935 (1985).

\bibitem{Kuzmin:1985mm}
  V.~A.~Kuzmin, V.~A.~Rubakov and M.~E.~Shaposhnikov,
  ``On the Anomalous Electroweak Baryon Number Nonconservation in the Early
  Universe,''
  Phys.\ Lett.\ B {\bf 155}, 36 (1985).

\bibitem{Bochkarev:1987wf}
  A.~I.~Bochkarev and M.~E.~Shaposhnikov,
  ``Electroweak Production of Baryon Asymmetry and Upper Bounds on the Higgs
  and Top Masses,''
  Mod.\ Phys.\ Lett.\ A {\bf 2}, 417 (1987).

\bibitem{Moore:1999fs}
  G.~D.~Moore and K.~Rummukainen,
  ``Classical Sphaleron Rate on Fine Lattices,''
  Phys.\ Rev.\ D {\bf 61}, 105008 (2000)
  [hep-ph/9906259].

\bibitem{Ibe:2004tg}
  M.~Ibe, R.~Kitano, H.~Murayama and T.~Yanagida,
  ``Viable Supersymmetry and Leptogenesis with Anomaly Mediation,''
  Phys.\ Rev.\ D {\bf 70}, 075012 (2004)
  [hep-ph/0403198].

\bibitem{Kolb:1979qa}
  E.~W.~Kolb and S.~Wolfram,
  ``Baryon Number Generation in the Early Universe,''
  Nucl.\ Phys.\ B {\bf 172}, 224 (1980)
  [Erratum-ibid.\ B {\bf 195}, 542 (1982)].

\end{thebibliography}
\end{document}